\documentclass[aps,prd,nofootinbib]{revtex4-1}
\usepackage{graphicx}				
\usepackage{epsfig}
\usepackage{bm}
\usepackage{amssymb}
\usepackage{amsmath}
\usepackage{color}			
\usepackage{amssymb}

\usepackage{hyperref}
\hypersetup{
  colorlinks   = true,    
  urlcolor     = blue,    
  linkcolor    = blue,    
  citecolor    = red      
}

\newcommand{\be}{\begin{equation}}
\newcommand{\ee}{\end{equation}}
\newcommand{\bea}{\begin{eqnarray}}
\newcommand{\eea}{\end{eqnarray}}

\newcommand{\dd}{{\rm d}}
\newcommand{\nn}{\nonumber}
\newcommand{\tr}{\mathrm{tr\, }}

\def\simge{\mathrel{%
   \rlap{\raise 0.511ex \hbox{$>$}}{\lower 0.511ex \hbox{$\sim$}}}}
\def\simle{\mathrel{
   \rlap{\raise 0.511ex \hbox{$<$}}{\lower 0.511ex \hbox{$\sim$}}}}

\begin{document}
\title{Sub-femtometer scale color charge correlations in the proton}

\author{Adrian Dumitru}
\email{adrian.dumitru@baruch.cuny.edu}
\affiliation{Department of Natural Sciences, Baruch College,
CUNY, 17 Lexington Avenue, New York, NY 10010, USA}
\affiliation{The Graduate School and University Center, The City
  University of New York, 365 Fifth Avenue, New York, NY 10016, USA}

\author{Vladimir Skokov}
\email{vskokov@ncsu.edu}
\affiliation{Department of Physics, North Carolina State University,
  Raleigh, NC 27695}
\affiliation{RIKEN BNL Research Center, Brookhaven National
  Laboratory, Upton, NY 11973, USA}

\author{Tomasz Stebel}
\email{tomasz.stebel@uj.edu.pl}
\affiliation{Institute of Nuclear Physics PAN, Radzikowskiego 152, 31-342 Krakow, Poland}
\affiliation{Institute of Physics, Jagiellonian University, S.\ Lojasiewicza 11, 30-348 Krakow, Poland}

\date{\today}

\begin{abstract}
  Color charge correlations in the proton at moderately small $x\sim
  0.1$ are extracted from its light-cone wave function.
  The charge fluctuations are far from Gaussian and they
  exhibit interesting dependence on impact parameter and on the
  relative transverse momentum (or distance) of the gluon probes. We
  provide initial conditions for small-$x$ Balitsky-Kovchegov
  evolution of the dipole scattering amplitude with impact parameter
  and $\hat r \cdot \hat b$ dependence, and with non-zero $C$-odd
  component due to three-gluon exchange. Lastly, we compute the
  (forward) Weizs\"acker-Williams gluon distributions, including the
  distribution of linearly polarized gluons, up to fourth order in
  $A^+$. The correction due to the quartic correlator provides a
  transverse momentum scale, $q\simge 0.5$~GeV, for nearly maximal
  polarization.
\end{abstract}

\maketitle

\section{Introduction}

The planned high luminosity electron ion collider (EIC) is designed to
perform ``imaging'' of the proton (and of heavy ions) with
unprecedented accuracy~\cite{EIC}. It will provide detailed
multi-dimensional parton distributions and insight into the
light-front wave function (LFwf) of the proton via high-energy
$\gamma^{(*)} - p$ scattering. The purpose of this paper is to expose
the color charge correlations obtained from the LFwf of the proton.\\

The concept of color charge density fluctuations in the transverse
impact parameter plane emerges naturally in high-energy (small-$x$)
scattering. The projectile charge traverses without recoil the (color)
field produced coherently by all ``valence'' charges in the target,
and its propagator is given by a path ordered exponential of that
field, c.f.\ sec.~\ref{sec:DipScattAmpl} below. For scattering of a
(virtual) photon from a proton target, this regime of coherent eikonal
scattering may set in at $x\simle 0.1$ where the longitudinal coherence
length $\sim 1/(x M_p)$ of the process in the rest frame of the proton
begins to exceed its radius. Nuclear targets, on the other hand, require $x
\simle 0.1/A^{1/3}$, where $A$ denotes the atomic number.

The scale separation in soft coherent fields sourced by random,
``frozen'' valence charges was introduced by McLerran and
Venugopalan~(MV) in ref.~\cite{MV}. Their model, devised for a very large
nucleus, describes Gaussian fluctuations of {\em classical} color
charge densities at vanishing momentum transfer: $\langle \rho^a(\vec
q_1) \, \rho^b(\vec q_2)\rangle_{\text{MV}} \sim \mu^2\, \delta(\vec
q_1 + \vec q_2)$.  However, when the density of valence charges in the
target is not very large, one would rather take the two-dimensional
color charge density as an operator acting on the light-front wave
function (LFwf) of the target~\cite{DMV}. We shall see that in the
regime of moderate $x\sim 0.1$ color charge fluctuations in
the proton are not Gaussian, and dependent on impact parameter and on
the transverse distance scale they are probed at.\\

After analyzing color charge correlations in the proton we proceed to
specify initial conditions for small-$x$ Balitsky-Kovchegov (BK)
evolution~\cite{BK} of the dipole scattering amplitude. Detailed fits
of BK evolution with running coupling corrections to the $\gamma^* -
p$ cross section measured at HERA have been performed by Albacete {\it
  et al.} in ref.~\cite{Albacete:2009fh}. More recent fits improve the
accuracy of the theory by employing a collinearly improved BK
evolution equation (ref.~\cite{Ducloue:2019jmy} and references
therein). However, such fits of small-$x$ QCD evolution to HERA DIS
data typically impose simplified, {\it ad-hoc} initial conditions for
the dipole scattering amplitude on the proton, starting at
$x=10^{-2}$. We attempt to construct initial conditions based on the
light-front wave function (LFwf) of the proton so that one may take
advantage of ``proton imaging'' performed at a future electron-ion
collider (EIC)~\cite{EIC}. We use a model LFwf to show that
interesting, non-trivial transverse momentum and impact parameter
dependent color charge correlations in the proton should be
expected. Furthermore, these initial conditions include a non-zero
$C$-odd ``Odderon'' contribution to the dipole scattering amplitude
which may be evolved to smaller $x$~\cite{Kovchegov:2003dm} in order
to address high-energy exclusive processes involving $C$-odd
exchanges;  or some spin dependent Transverse Momentum Dependent (TMD)
distributions such as the (dipole) gluon Sivers
function of a transversely polarized proton~\cite{Yao:2018vcg}. \\

Our final objective is to compute the Weizs\"acker-Williams (forward)
gluon distributions, in particular the distribution of linearly
polarized gluons, at next-to-leading (fourth) order in $A^+$
(sec.~\ref{sec:WW}). At this order the conventional and linearly
polarized distributions no longer coincide, and they involve the
correlator of four color charge density operators in the proton. This
is an independent correlation function which can not be reduced to
products of quadratic color charge correlators like in an effective
theory of Gaussian color charge fluctuations.  The WW gluon
distribution is a TMD, its general operator definition has been
provided in refs.~\cite{WWoperator}. The WW gluon TMDs appear in a
variety of processes such as production of a dijet or heavy quark pair
in hadronic collisions~\cite{WWhadronic} or DIS at
moderate~\cite{WW-DISmoderate} or high
energies~\cite{Dominguez:2011br,Dumitru:2015gaa,WW-dijet-smallx};
photoproduction of three jets~\cite{Altinoluk:2020qet}; photon
pair~\cite{WWphoton}, quarkonium~\cite{WW-quarkonium}, quarkonium
pair~\cite{WW-quarkonium-pair}, or quarkonium plus
dilepton~\cite{WW-quarkonium-dilepton} production in hadronic
collisions. These gluon distributions also determine the fluctuations
of the divergence of the Chern-Simons current at the initial time of a
relativistic heavy-ion collision~\cite{WW-CS}.

\section{Setup}  \label{sec:LFwf}
The light cone state of an unpolarized on-shell proton with
four-momentum $P^\mu = (P^+, P^-,\vec{P}_\perp)$ is written
as~\cite{Lepage:1980fj}
\bea
|P\rangle &=& \frac{1}{\sqrt{6}} \int \frac{\dd x_1\dd x_2 \dd x_3}
{\sqrt{x_1 x_2 x_3}} 
\delta(1-x_1-x_2-x_3)
\int \frac{\dd^2 k_1 \dd^2 k_2 \dd^2 k_3}{(16\pi^3)^3}\,
 16\pi^3 \delta(\vec{k}_1+\vec{k}_2+\vec{k}_3)\nonumber\\
  &\times& 
 \psi(x_1, \vec k_1; x_2, \vec k_2; x_3, \vec k_3)
 \sum_{i_1, i_2, i_3}\epsilon_{i_1 i_2 i_3}
  |p_1,i_1; \, p_2,i_2; \, p_3,i_3\rangle~.  \label{eq:def_|P>}
\label{eq:valence-proton}
\eea
The $n$-parton Fock space amplitudes are universal and process
independent. They encode the non-perturbative structure of hadrons.
Here, we have restricted to the valence quark Fock state, assuming
that the process probes parton momentum fractions of order $x\sim
0.1$, and moderately high transverse momenta. In this regime, the
above should be a reasonable first approximation.

The three on-shell quark momenta are specified by their lightcone
momentum components $p_i^+ = x_i P^+$ and their transverse components
$\vec{p}_{i} = x_i \vec{P}_\perp + \vec{k}_i$. The colors of the
quarks are denoted by $i_{1,2,3}$.  We omit helicity quantum numbers
(and flavor indices) as they play no role in our analysis. $\psi$ is
symmetric under exchange of any two of the quarks, and is normalized
according to
\be \label{eq:Norm_psi3}
 \int {\dd x_1\dd x_2 \dd x_3}\, \delta(1-x_1-x_2-x_3)
  \int \frac{{\dd^2 k_1 \dd^2 k_2 \dd^2 k_3}}{(16\pi^3)^3}\,
  (16\pi^3)\,\delta(\vec{k}_1+\vec{k}_2+\vec{k}_3)\, 
  |\psi|^2 = 1~.
\ee
This corresponds to the proton state normalization
\bea
\langle K | P\rangle &=& 16\pi^3 \, P^+ \delta(P^+ - K^+)
\, \delta(\vec{P}_\perp - \vec{K}_\perp) \label{eq:ProtonNorm1}
~.
\eea
Below, we neglect plus momentum transfer so that $\xi = (K^+ -
P^+)/P^+ \to 0$. This approximation is valid at high energies.\\

For numerical estimates we employ a  model wave function
$\psi(x_1, \vec k_1; x_2, \vec k_2; x_3, \vec k_3)$ described in
appendix~\ref{app:BrodskySchlumpf}.

\section{Dipole scattering amplitude}  \label{sec:DipScattAmpl}

The $S$-matrix for scattering of a quark - antiquark dipole
off the fields in the target proton can be expressed as (see,
e.g.\ ref.~\cite{Mueller:2001fv})
\be \label{eq:S_dipole_b}
   {\cal S} (\vec r,\vec b) =
   \frac{1}{N_c}\,{\rm tr} \,\left< U\left(\vec b +
     \frac{\vec r}{2}\right)\,
     U^\dagger\left( \vec b - \frac{\vec r}{2}\right)\right> \, .
\ee
Following the standard convention in the small-$x$ literature we
define the scattering amplitude
\be  \label{eq:T_dipole_b}
{\cal T} (\vec r,\vec b) = 1-{\cal S} (\vec r,\vec b)~,
\ee
without a factor of $i$.

\begin{figure}[htb]
  \centering
  \includegraphics[width=0.45\textwidth]{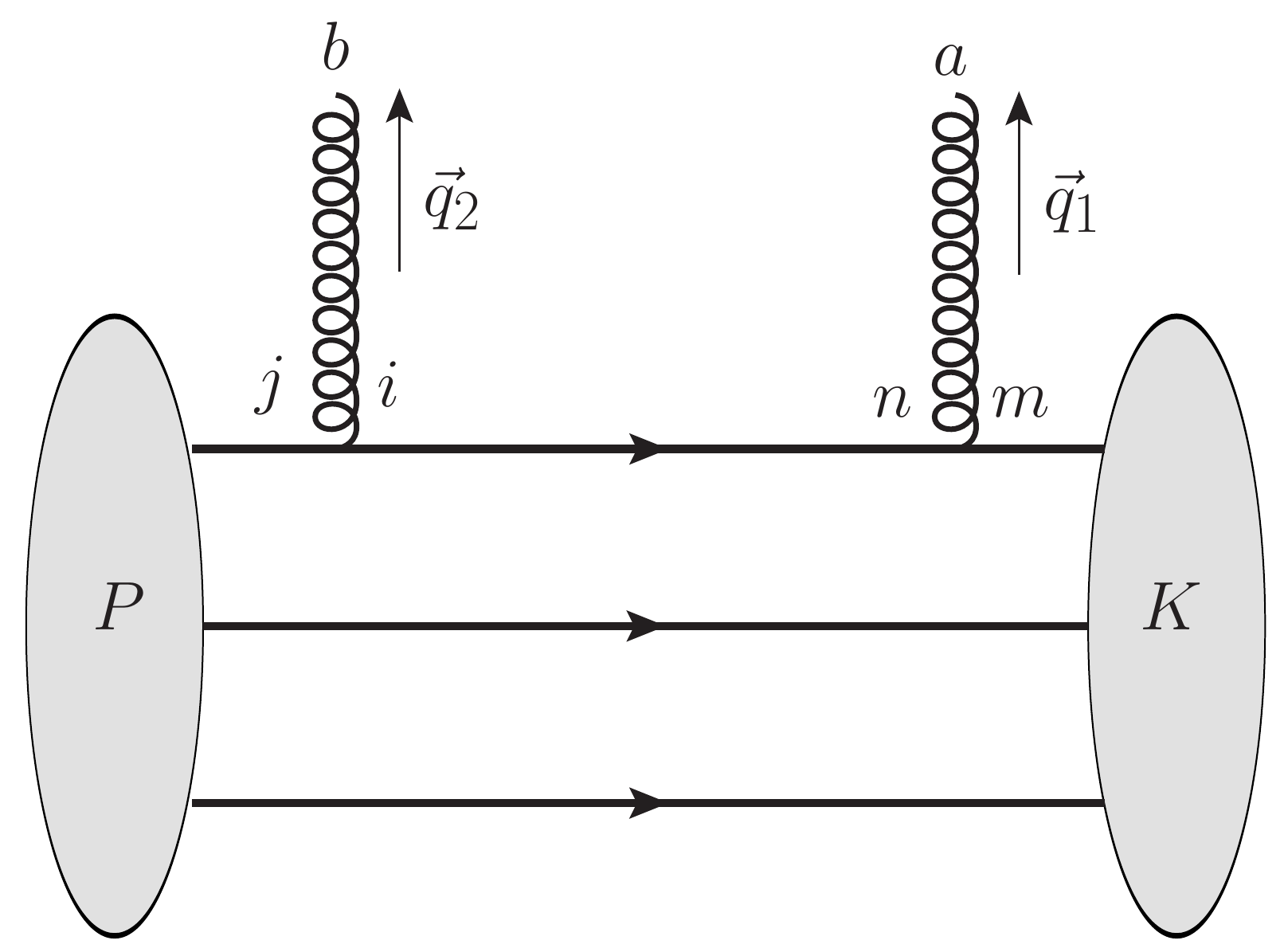}
  \hspace*{1.5cm}
  \includegraphics[width=0.45\textwidth]{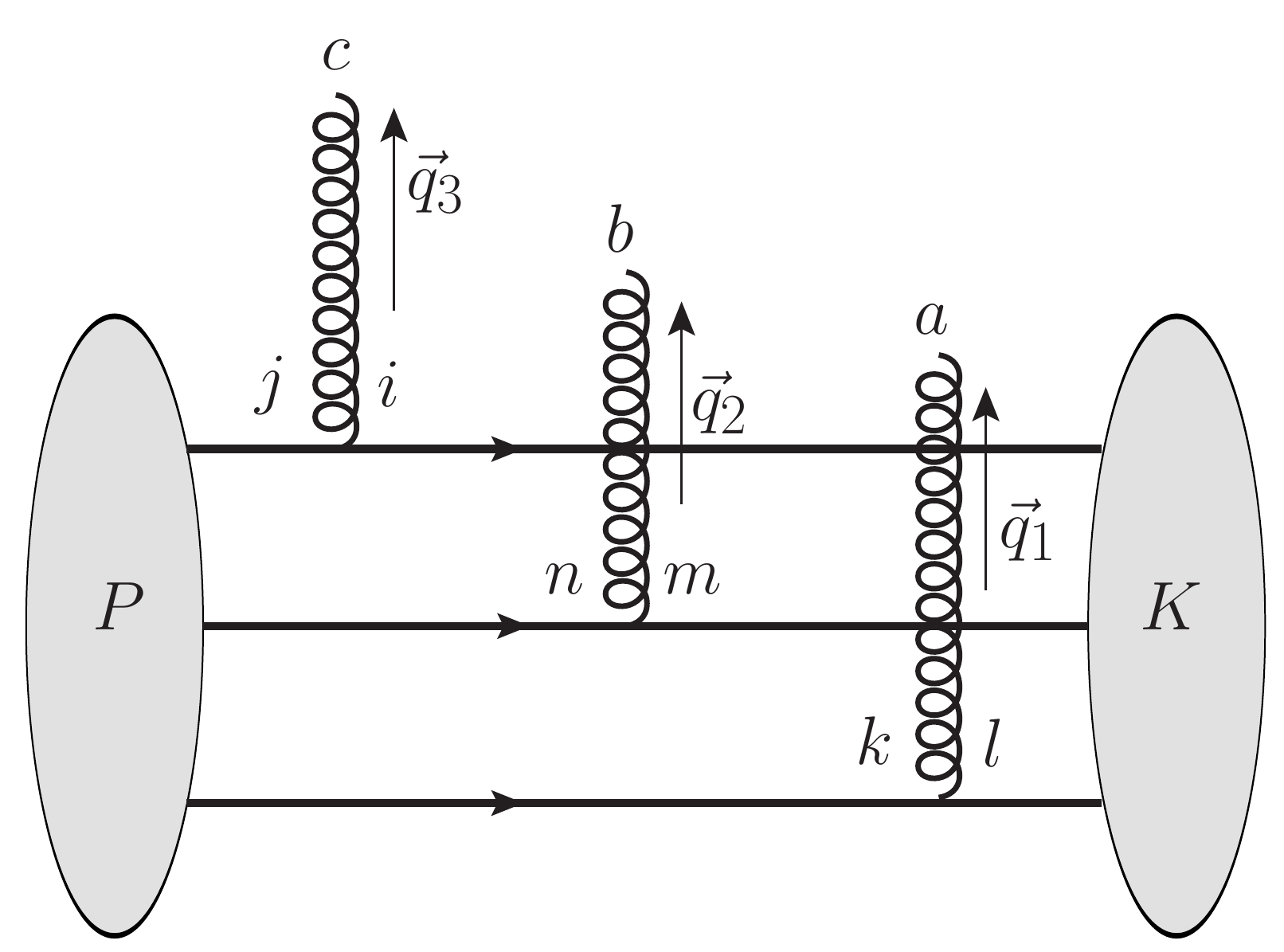}
  \caption{Left: one of the diagrams for the correlator $\langle
    \rho^a(\vec q_1)\, \rho^b(\vec q_2)\rangle$ (once Coulomb
    propagators are amputated); this contribution dominates at large
    relative gluon momenta but small total momentum transfer $\vec
    K_T=-\vec q_1 - \vec q_2$.\\ Right: one of the diagrams for the
    correlator $\langle \rho^a(\vec q_1)\, \rho^b(\vec q_2)\,
    \rho^c(\vec q_3)\rangle$; this contribution dominates when the
    three gluons share a large
    momentum transfer, $\vec K_T/3 \simeq -\vec q_1 \simeq - \vec q_2
    \simeq- \vec q_3$.}
\label{fig:diag-rhorho}
\end{figure}

When integrated over impact parameters $\vec b$,
eq.~(\ref{eq:T_dipole_b}) is related to the so-called dipole gluon
distribution~\cite{Dominguez:2011wm}.  Here, $U$ ($U^\dagger$) are
(anti-)path ordered Wilson lines representing the eikonal scattering
of the dipole of size $\vec r$ at impact parameter $\vec b$:
\be \label{eq:WilsonLines}
U(\vec x_T) = {\cal P} e^{ig \int dx^- A^{+a}(x^-,\vec x_T)\, t^a} ~~~~~,~~~~~
U^\dagger(\vec x_T) = \overline{\cal P} e^{-ig \int dx^- A^{+a}(x^-,\vec x_T)\, t^a} ~.
\ee
${\cal S} (\vec r,\vec b)$ and ${\cal T} (\vec r,\vec b)$ are
invariant under the simultaneous ${\cal P} \leftrightarrow
\overline{\cal P}$, $\vec r \to - \vec r$, $gA^+\to -gA^+$. We now
expand ${\cal T} (\vec r,\vec b)$ to third order in $gA^+$, neglecting
exchanges of more than three gluons, and write it in terms of
correlators of the field {\em integrated} over the longitudinal
coordinate:
\bea
A^{+a}(\vec x_T) &=& \int \dd x^- A^{+a}(\vec x_T, x^-)~, \nn\\
A^{+a}(\vec x_T) \, A^{+b}(\vec y_T) + A^{+b}(\vec y_T) \, A^{+a}(\vec x_T) &=&
{\cal P} \int \dd x^- \int \dd y^- A^{+a}(\vec x_T, x^-) A^{+b}(\vec
y_T, y^-) \nn\\
& & +
\overline{\cal P}
\int \dd x^- \int \dd y^- A^{+a}(\vec x_T, x^-) A^{+b}(\vec y_T, y^-)~.
\eea
This field is related to the 2d color charge density through
\be
- \nabla_\perp^2 A^{+a}(\vec x_T) = \rho^a(\vec x_T)~,
\ee
allowing us to express the dipole scattering amplitude in terms of
color charge density correlators. Some of the diagrams that contribute
to the two- and three-gluon exchange amplitudes are shown in
fig.~\ref{fig:diag-rhorho}. The general relation of
correlators of Wilson lines at small $x$ to Generalized Parton
Distributions has been elucidated in ref.~\cite{Altinoluk:2019wyu}, to
all twists.
\\

\begin{figure}[htb]
  \centering
  \includegraphics[width=0.48\textwidth]{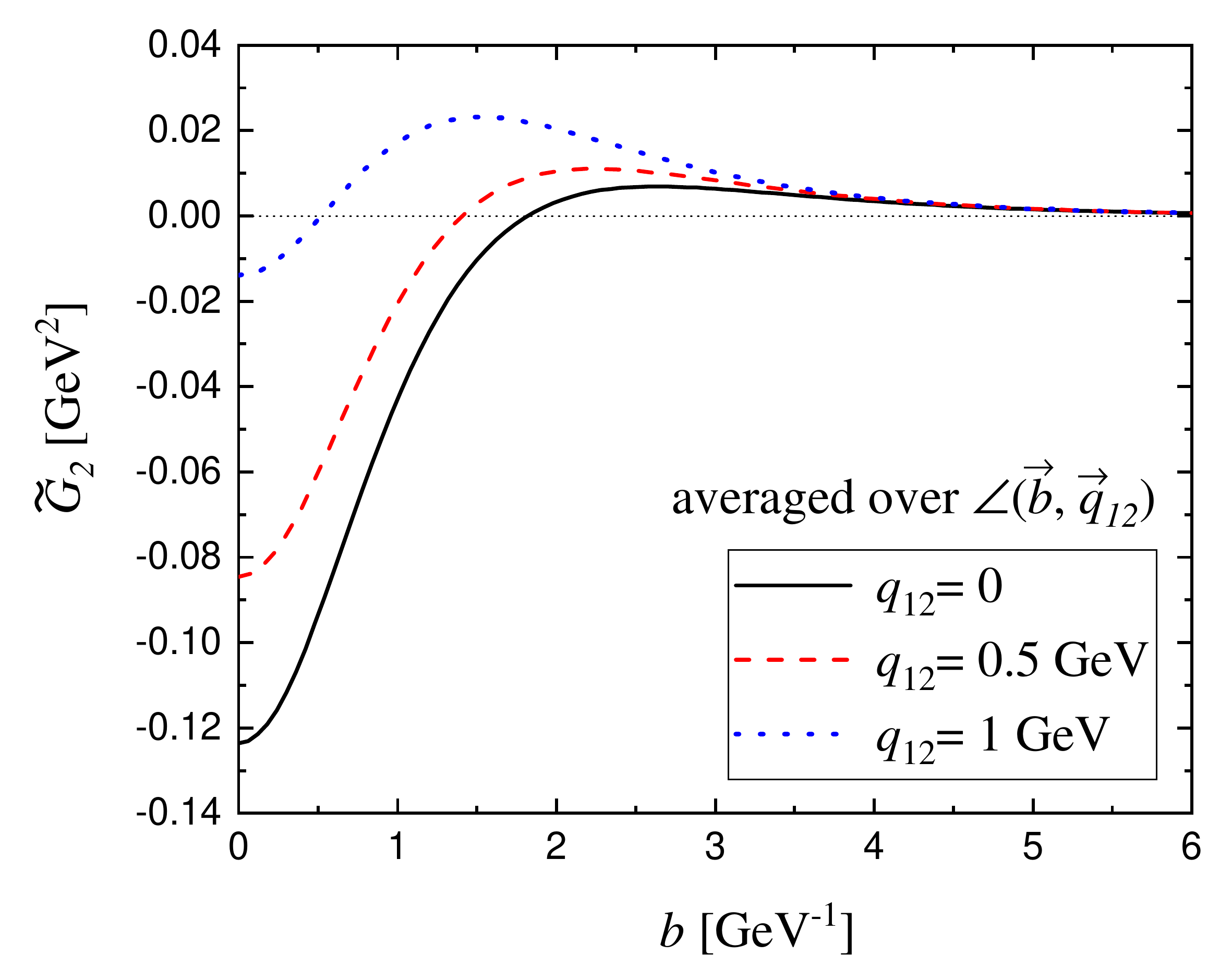}
  \includegraphics[width=0.48\textwidth]{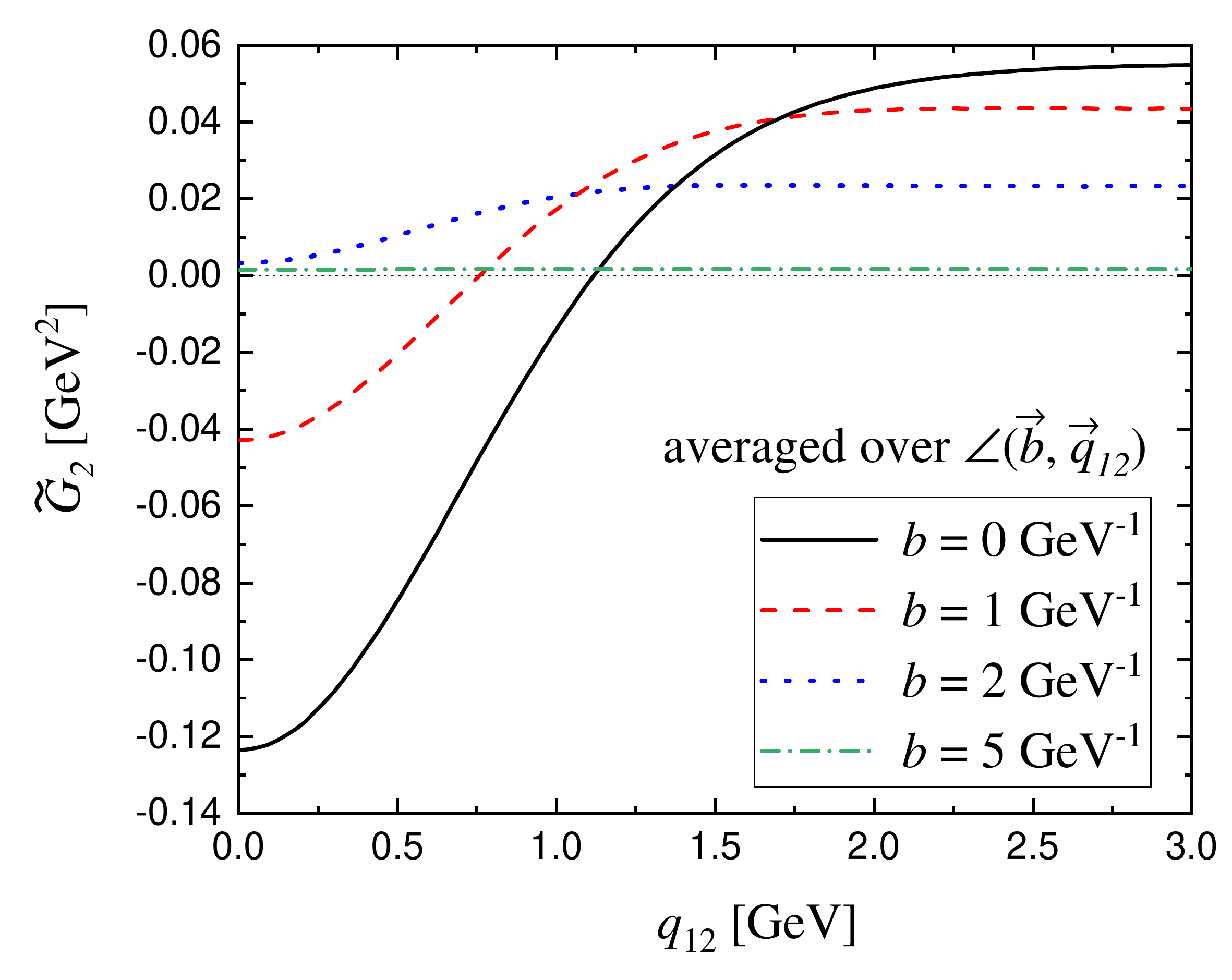}
  \caption{The quadratic color charge density correlator $\widetilde G_2(\vec
    q_{12},b)$ in the proton as a function of impact parameter and
    relative transverse momentum of the two gluon probes.}
\label{fig:G2_bq}
\end{figure}

$C$-even two gluon exchange corresponds to the scattering
amplitude~\cite{DMV}
\bea
{\cal T}_{gg}(\vec r,\vec b) &=& 
  - \frac{g^4}{2} C_F\,\int\limits_{K_T, q}
  \frac{e^{-i\vec b \cdot \vec K_T}}{(\vec q -\frac{1}{2}\vec K_T)^2
    \, (\vec q+ \frac{1}{2} \vec
  K_T)^2}\,\left(
\cos\left(\vec r \cdot {\vec q}\right) 
- \cos\left(\frac{{\vec r}\cdot \vec K_T}{2}\right)\right)
\,
G_2\left({\vec q}-\frac{1}{2}\vec K_T,-{\vec q}-\frac{1}{2}\vec K_T\right)
~.    \label{eq:Pomeron}
\eea
(We use the shorthand notation $\int_q = \int\dd^2q/(2\pi)^2$.)
Here, we introduced the color charge correlator
\be
\left< \rho^a(\vec q_1) \,
\rho^b(\vec q_2)\right> \equiv {\rm tr} \,t^a t^b\, g^2\, G_2(\vec q_1,\vec
q_2)~,
\ee
see appendix~\ref{sec:rho_Correlators} for details. It is symmetric
under a simultaneous sign flip of both arguments and so ${\cal
  T}_{gg}(\vec r,\vec b)$ is real. The integral in
eq.~(\ref{eq:Pomeron}) is free of infrared divergences since $G_2$
satisfies a Ward identity and vanishes when either one of the gluon
momenta goes to zero~\cite{Bartels:1999aw,Ewerz:2001fb}: $G_2\left({\vec
  q}-\frac{1}{2}\vec K_T,-{\vec q}-\frac{1}{2}\vec K_T\right) \sim
({\vec q}\pm\frac{1}{2}\vec K_T)^2$ as ${\vec q}\to\pm \frac{1}{2}\vec
K_T$. In fig.~\ref{fig:G2_bq} we show a numerical estimate for $G_2$
as a function of impact parameter $b$ or relative momentum $\vec
q_{12} = \vec q_1 - \vec q_2 = 2\vec q_1 + \vec K_T$:
\be
\widetilde G_2(\vec q_{12},\vec b) = \int_{K_T} e^{-i\vec b\cdot \vec K_T}\,
G_2\left(\frac{\vec q_{12}-\vec K_T}{2}, -\frac{\vec q_{12}+\vec
  K_T}{2} \right)~.
\ee
We also average over the relative directions of $\vec q_{12}$ and
$\vec b$. For numerical estimates we used the model wave function by
Brodsky and Schlumpf~\cite{Brodsky:1994fz} described briefly in
appendix~\ref{app:BrodskySchlumpf}.

$G_2$ measures charge correlations seen by two gluon probes of the
same color. There is a color charge anti-correlation (``repulsion'')
at small relative momentum of the gluon probes in the center of the
proton which turns into a positive correlation (``attraction'')
towards the periphery, or at high relative momentum. The integral of
$\widetilde G_2$ over the 2d impact parameter plane at vanishing relative
momentum is zero:
\be
\int \dd^2b\,\, \widetilde G_2(\vec q_{12}=0,\vec b) = 0~.
\ee
A similar relation holds for the cubic charge correlators discussed below.
\\

At third order in $A^{+a}$ we have the following scattering amplitude
for $C$-odd three gluon exchange~\cite{DMV}:
\bea
{\cal T}_{ggg}(\vec r,\vec b) &=&  
\frac{5}{18}\, g^6
\int\limits_{q_1, q_2, q_3}
\frac{1}{q_1^2}\frac{1}{q_2^2}\frac{1}{q_3^2}\,
e^{-i \vec b \cdot \vec K_T}\,
G_3^-(\vec q_1,\vec q_2,\vec q_3)\,\left[
\sin\left(\vec r\cdot \vec q_1 + \frac{1}{2} \vec r\cdot \vec
    K_T\right)
    - \frac{1}{3}\sin\left(\frac{{1}}{2}\vec r\cdot
  \vec K_T\right)\right]~.
\label{eq:Odderon-operator}
\eea
Here, $\vec K_T \equiv - (\vec q_1 + \vec q_2 + \vec q_3)$. We
denote the $C$-odd part of the correlator of three color
charges as
\be
\left<
\rho^a(\vec q_1) \, \rho^b(\vec q_2)\, \rho^c(\vec q_3)\right>_{C=-} \equiv
\frac{1}{4} d^{abc}\, g^3\, G_3^-(\vec q_1,\vec q_2,\vec q_3)
\ee
This
correlator, too, is symmetric under a simultaneous sign flip of
all three gluon momenta and so ${\cal T}_{ggg}(\vec r,\vec b)$ is imaginary.
Also, it vanishes quadratically in any of the transverse momentum
arguments so that ${\cal T}_{ggg}(\vec r,\vec b)$ is free of infrared
divergences.

\begin{figure}[htb]
  \centering
  \includegraphics[width=0.48\textwidth]{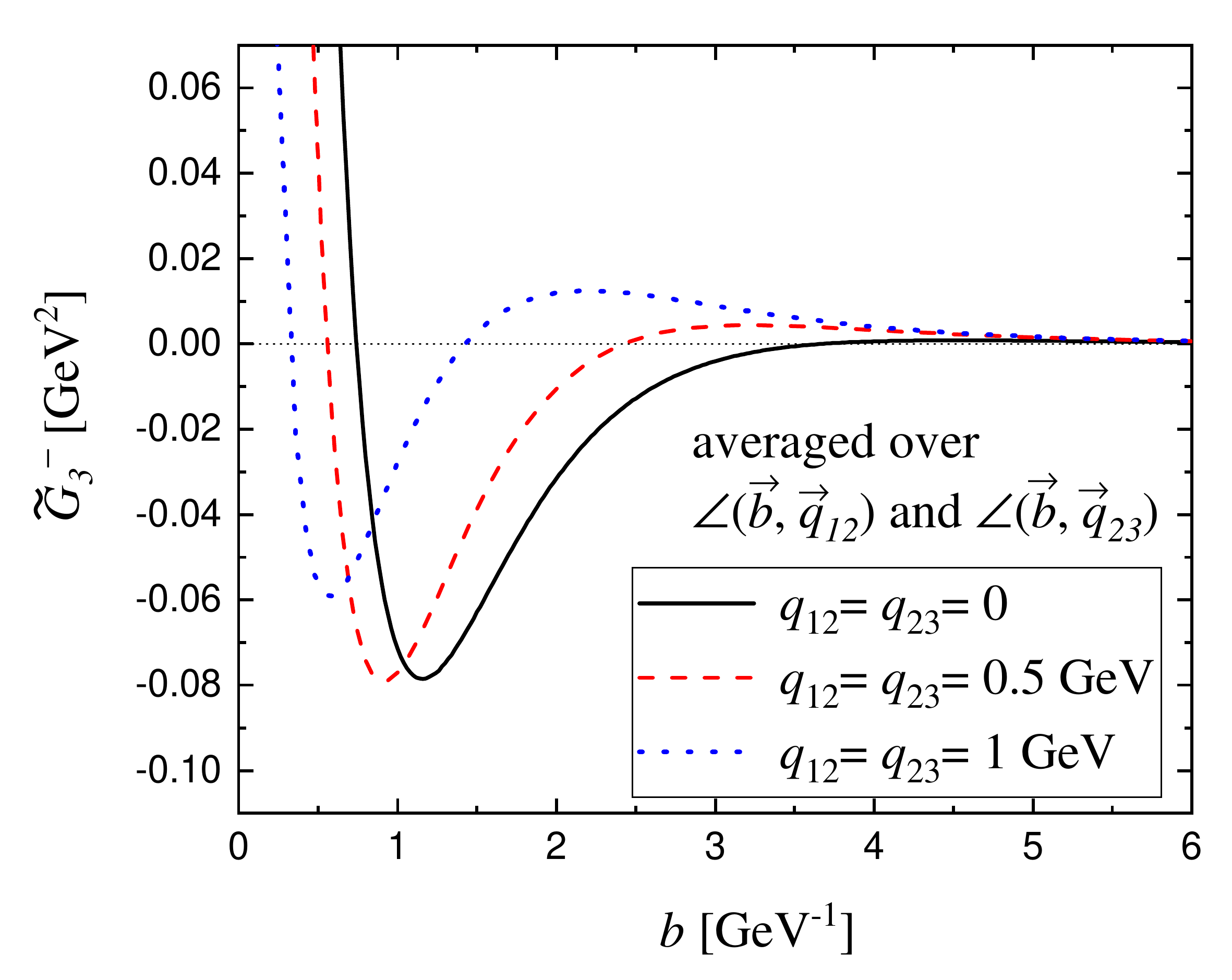}
  \includegraphics[width=0.48\textwidth]{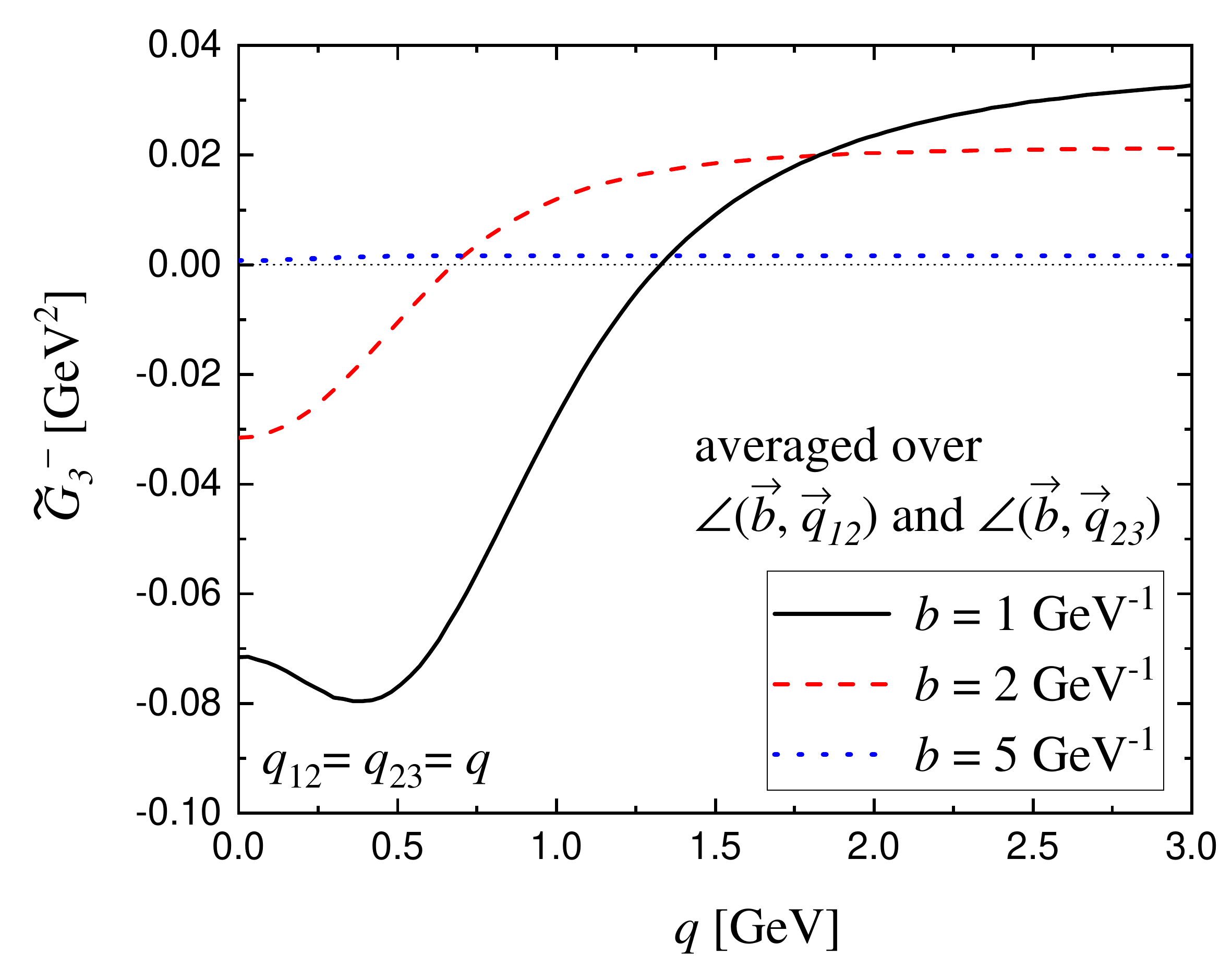}
  \caption{The $C$-odd part of the cubic color charge density
    correlator $\widetilde G_3^-$ in the proton as a function of
    impact parameter and relative transverse momentum.}
\label{fig:G3-_bq}
\end{figure}
The fact that $G_3^-$ does not vanish shows that color charge
fluctuations in the proton state~(\ref{eq:valence-proton}) are not
Gaussian.  A numerical estimate of $\widetilde G_3^-$ is shown in
fig.~\ref{fig:G3-_bq}. At small relative momentum we observe a
positive correlation at the center of the proton; $\widetilde
G_3^-(b)$ diverges logarithmically at $b\to0$ due to contributions
from large momentum transfer $-t=K_T^2$. This turns into an
anti-correlation around $b\approx 1$~GeV$^{-1}$, and then vanishes for
large impact parameters. At high relative momentum the correlator is
large and positive at small $b$. For generic impact parameters and
momenta $\widetilde G_2$ and $\widetilde G_3^-$ are of similar
numerical magnitude.\\

For completeness, we finally show the $C$-even part of the
correlator of three color charges,
\be
\left<
\rho^a(\vec q_1) \, \rho^b(\vec q_2)\, \rho^c(\vec q_3)\right>_{C=+} \equiv
\frac{i}{4} f^{abc}\, g^3\, G_3^+(\vec q_1,\vec q_2,\vec q_3)~,
\ee
even though it does not contribute to the dipole scattering amplitude.
\begin{figure}[htb]
  \centering
  \includegraphics[width=0.48\textwidth]{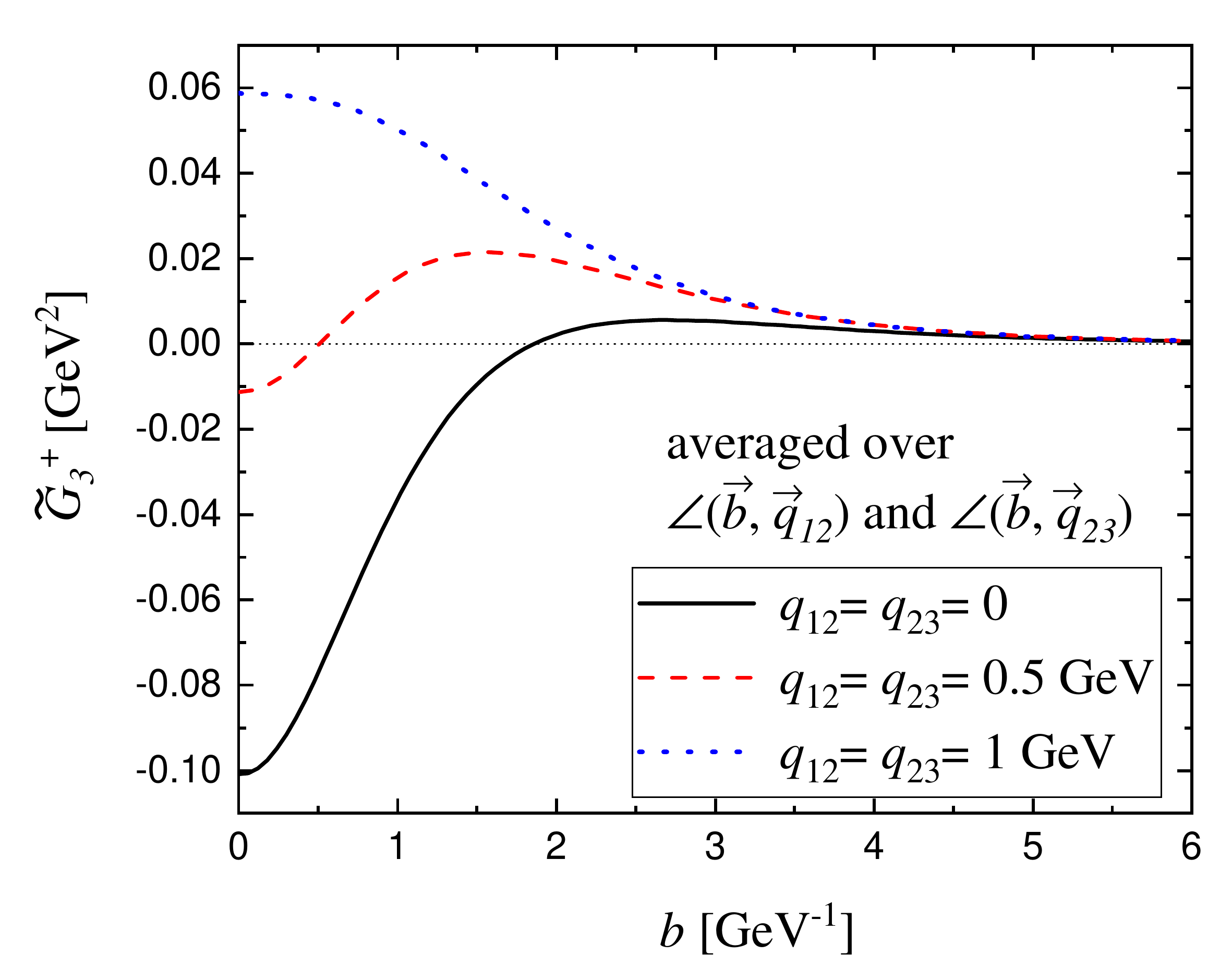}
  \includegraphics[width=0.48\textwidth]{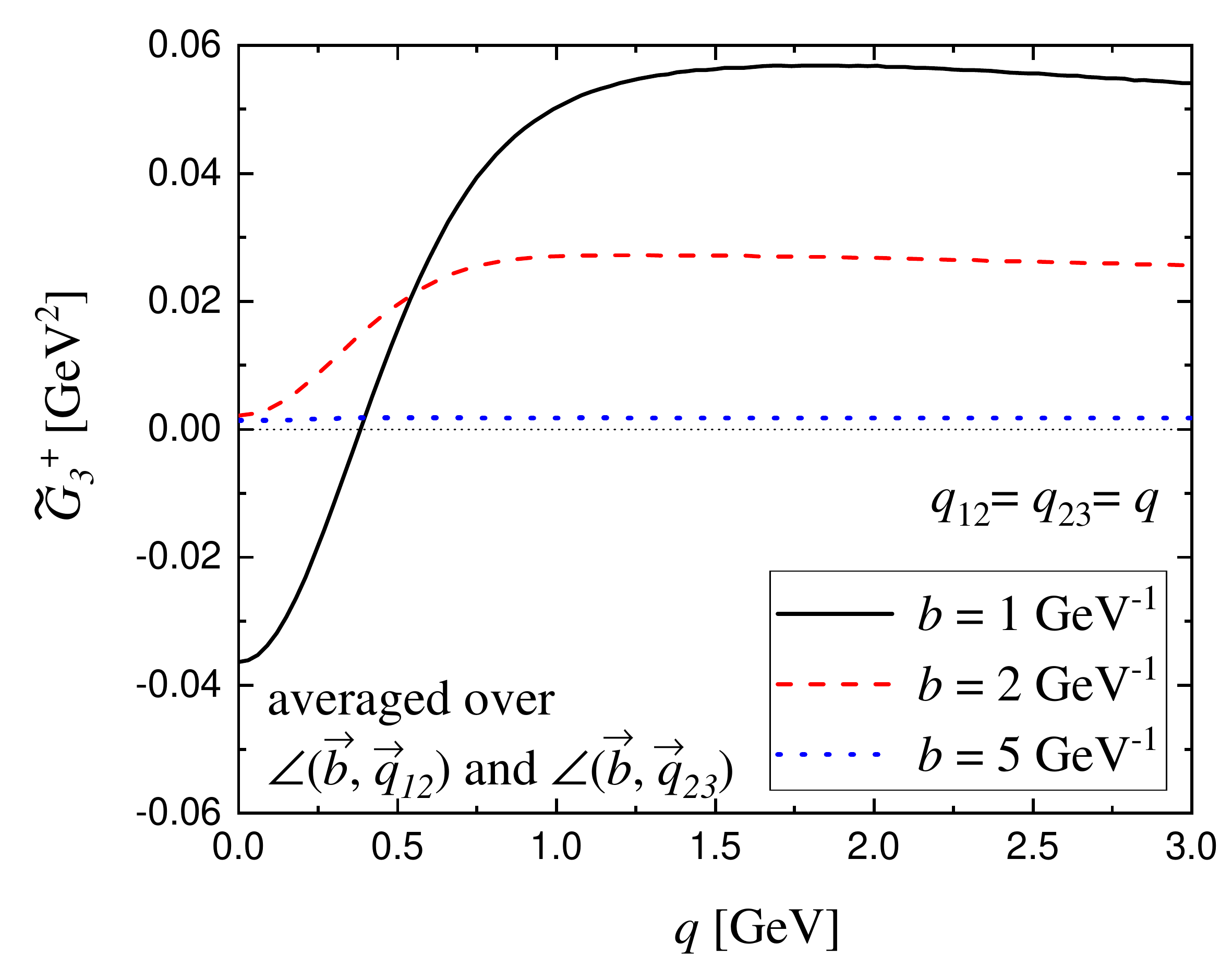}
  \caption{The $C$-even part of the cubic color charge density
    correlator $\widetilde G_3^+$ in the proton as a function of
    impact parameter and relative transverse momentum.}
\label{fig:G3+_bq}
\end{figure}
This correlator is negative near the center, and for small relative
momenta, then turns into a positive correlation at large momenta.\\

All three color charge correlators decay with increasing impact
parameter, just as expected intuitively. Observing the correlations at
small $b$ involves large momentum transfer to the proton to zoom in on
its center.  The regime where the exchanged gluons share a large
momentum transfer $-t = K_T^2$ is dominated by $n$-body diagrams such
as the one shown in fig.~\ref{fig:diag-rhorho}(right), where the
static gluons attach to as many sources as possible\footnote{This was
  first noted by Donnachie and Landshoff who argued that three gluon
  exchange should dominate over two-gluon exchange in elastic
  proton-proton scattering at high energy and large $-t$ ($\ll
  s$)~\cite{Donnachie:1979yu}.}~\cite{Dumitru:2019qec}. This leads to
the greatest overlap of the wave functions of incoming and scattered
proton.\\

We now show the behavior of the dipole scattering amplitude ${\cal
  T}(\vec b, \vec r)$. For all figures we assumed a fixed $\alpha_s =
0.35$~\cite{Dumitru:2019qec,Low:1975sv} and we align the impact parameter and
dipole vectors. However, the scattering amplitude does depend on the
relative orientation of $\vec b$ and $\vec r$~\footnote{This would
  give rise to azimuthal correlations in double parton scattering in
  hadronic collisions~\cite{Hagiwara:2017ofm}.}.

\begin{figure}[htb]
  \centering
  \includegraphics[width=0.45\textwidth]{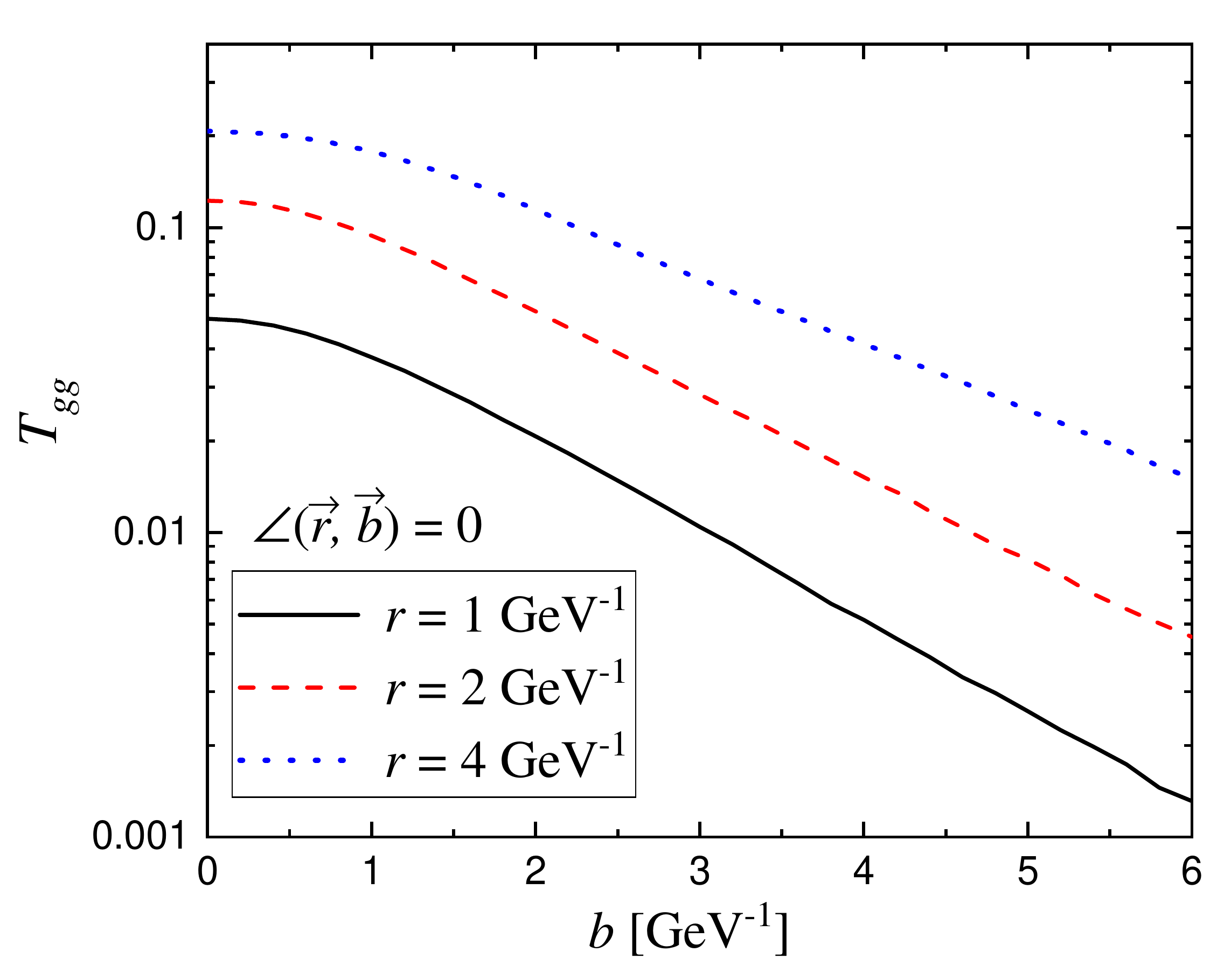}
  \includegraphics[width=0.45\textwidth]{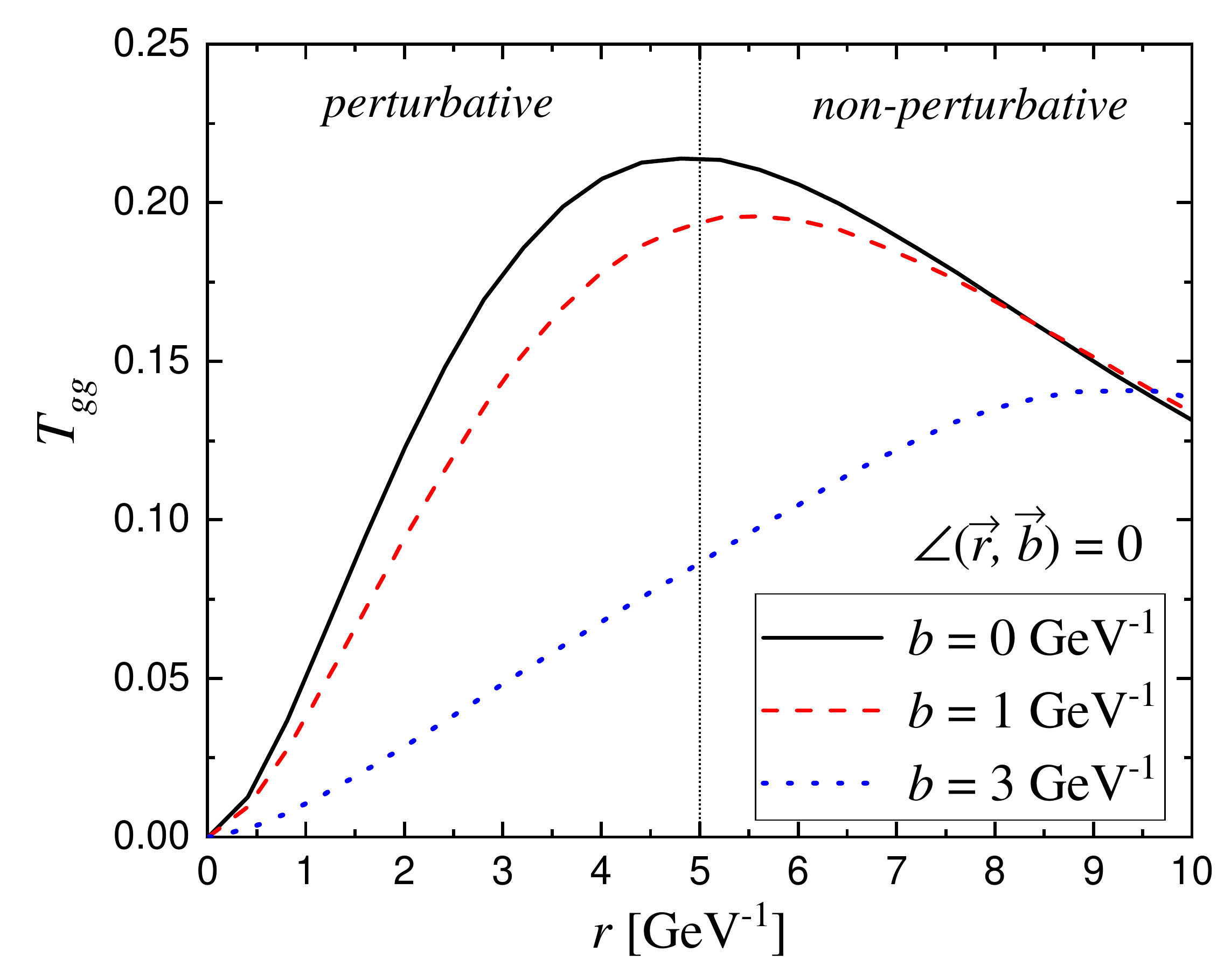}
  \caption{The two gluon exchange amplitude ${\cal T}_{gg}(\vec
b, \vec r)$.}
\label{fig:Tgg}
\end{figure}
The two gluon exchange amplitude ${\cal T}_{gg}(\vec b, \vec r)$ is
shown in fig.~\ref{fig:Tgg}. It displays the expected roughly
exponential falloff at large impact parameters. The amplitude is
significantly smaller than 1 even at the center of the proton,
albeit not by several orders of magnitude, e.g.\ ${\cal T}_{gg}
\simeq 0.1$ at $b=1$~GeV$^{-1}$ and $r=2$~GeV$^{-1}$. Matching this to
${\cal T}_{gg} = \frac{1}{4} r^2 Q_s^2(b)$ would correspond to a
saturation momentum of about $Q_s(b) \approx 0.3$~GeV at
$b=1$~GeV$^{-1}$ and $x\sim0.1$. For comparison, we recall $Q_s
\approx 0.4-0.5$~GeV at $x=0.01$, on average over impact parameters,
extracted from systematic fits of BK evolution with running coupling
corrections to HERA data for $F_2$~\cite{Albacete:2009fh}.

As expected, ${\cal T}_{gg}(\vec r)$ at fixed $b$ first increases with
the size of the dipole; the slope is less steep at larger impact
parameters where the target is more ``dilute''.  The scattering
amplitude eventually reaches a maximum value for $r_{\text{max}}
\simge 5$~GeV$^{-1}$ beyond which it decreases again as the projectile
dipole ``misses'' the target\footnote{This behavior also emerges as a
  consequence of impact parameter dependent small-$x$ BK evolution,
  even when the dipole amplitude at the initial $x_0$ increases
  monotonically with $r$~\cite{GolecBiernat:2003ym}.}.  However, this
behavior occurs in a regime of large dipoles where the analysis of the
scattering amplitude (and of $\gamma^{(*)} \to q\overline{q}$) in
perturbation theory is not valid.

\begin{figure}[htb]
  \centering
  \includegraphics[width=0.45\textwidth]{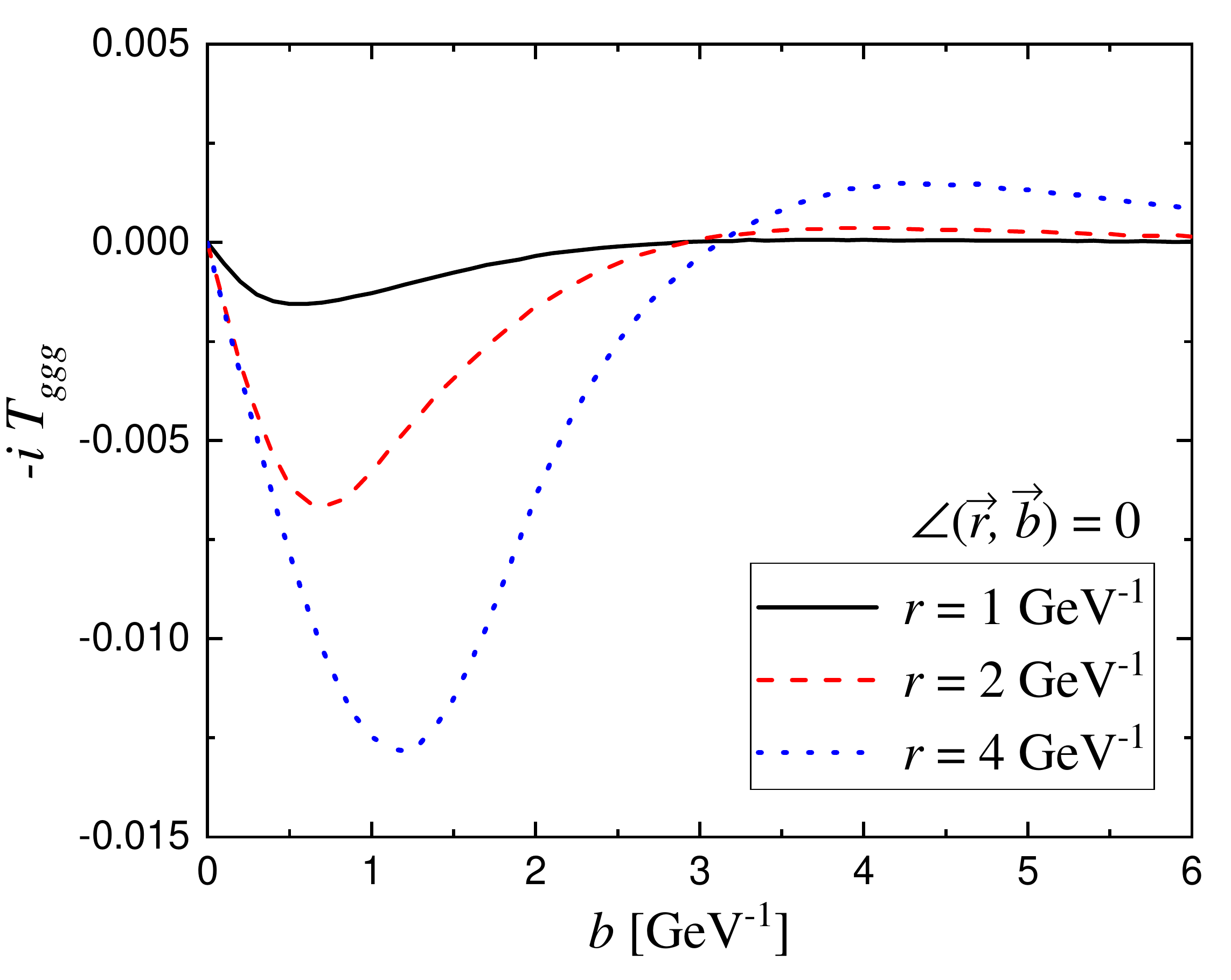}
  \includegraphics[width=0.45\textwidth]{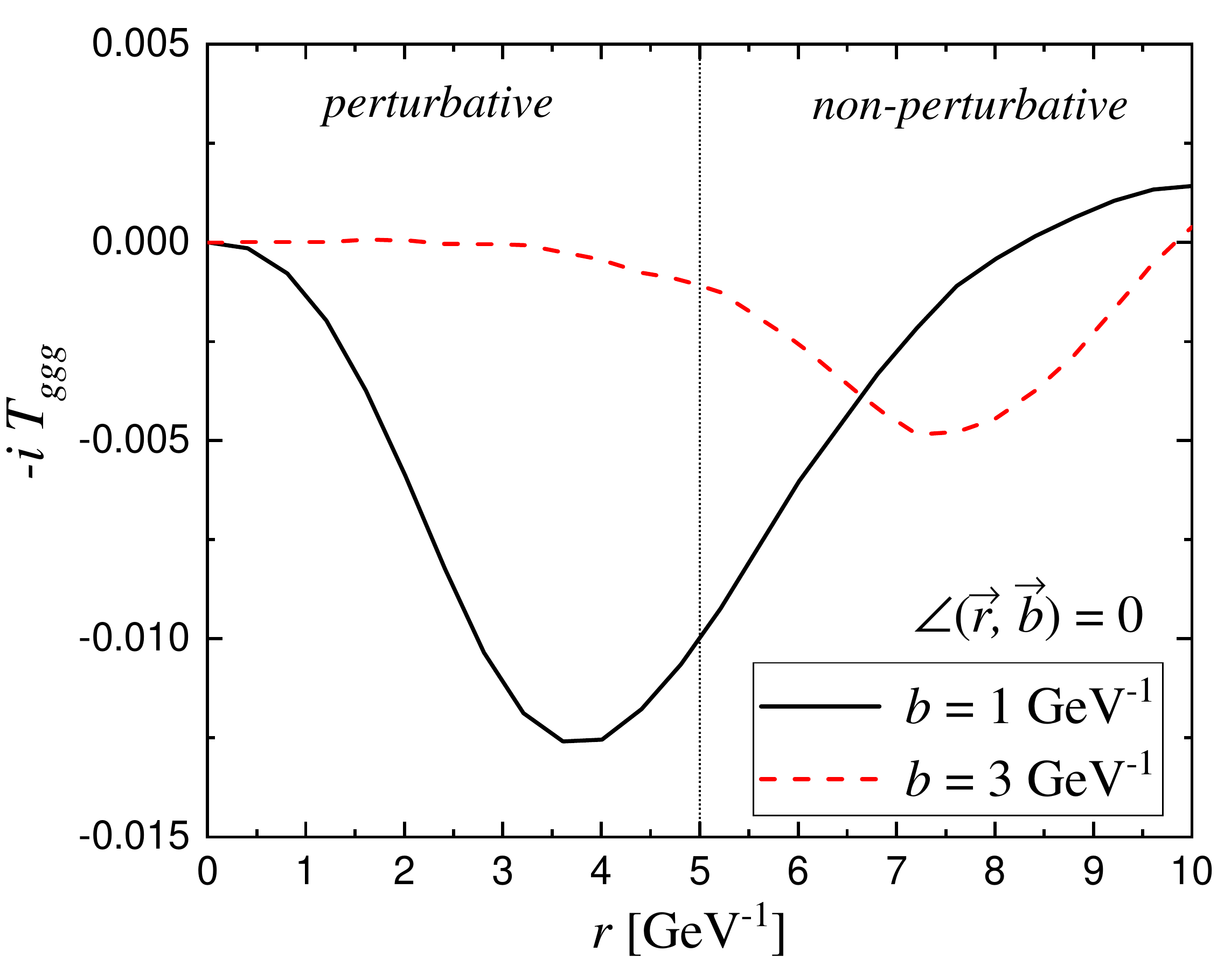}
  \caption{The $C$-odd three gluon exchange amplitude Im~${\cal
      T}_{ggg}(\vec b, \vec r)$.}
\label{fig:Tggg}
\end{figure}
The $C$-odd three gluon exchange amplitude (``Odderon''\footnote{We
  should mention that we restrict to the Odderon associated with
  (relatively large) transverse momentum transfer $\vec K_T$. For
  nearly forward scattering another Odderon exchange associated with a
  spin flip of the proton may appear~\cite{spinOdderon}. }) $-i{\cal
  T}_{ggg}(\vec b, \vec r)$ is shown in fig.~\ref{fig:Tggg}. This
amplitude changes sign under $\vec b \to - \vec b$ (negative parity)
and vanishes at $b=0$. Its magnitude is maximal at $b\sim 0.5 -
1.2$~GeV$^{-1}$, approximately where the gradient of the two-gluon
exchange amplitude is greatest~\cite{KovSievert}. For impact
parameters $b\simle 3$~GeV$^{-1}$ and small dipoles, $r \simle
4$~GeV$^{-1}$, we find that ${\cal T}_{ggg}$ is smaller than ${\cal
  T}_{gg}$ by at least one order of magnitude\footnote{The magnitude
  of Im~${\cal T}_{ggg}$ obtained from the present LFwf is one order of
  magnitude smaller than the one used as the initial condition for
  small-$x$ evolution in ref.~\cite{Yao:2018vcg}, where the authors
  compute the dipole gluon Sivers function in a transversely polarized
  proton.}. This is not because color charge fluctuations in the
proton are nearly Gaussian, as the magnitudes of $G_2$ and $G_3^-$
(shown above) are similar. Rather, it appears to originate mostly from
the parity odd nature of ${\cal T}_{ggg}$ which gives rise to large
cancellations in the integral in eq.~(\ref{eq:Odderon-operator}). As a
consequence, semi-hard processes requiring $C$-odd three gluon
exchange have small
cross-sections~\cite{Dumitru:2019qec}. Alternatively, one may search
for the perturbative Odderon via charge asymmetries in diffractive
electroproduction of a $\pi^+\, \pi^-$ pair~\cite{Hagler:2002nh}.

\section{Weizs\"acker-Williams gluon distributions} \label{sec:WW}

In this section we relate the color charge correlators to the
(forward) WW gluon distribution. It is given, at small-$x$, by the
correlator of two light-cone gauge fields~\cite{Dominguez:2011wm,WW}
\be \label{eq:WW-Gij}
xG^{ij}_{\text{WW}}(x,\vec q) = \frac{1}{2}\delta^{ij}\,
xG^{(1)}(x,\vec q) + \frac{1}{2}\left(2\frac{q^i
  q^j}{q^2}-\delta^{ij}\right)\, xh_\perp^{(1)}(x,\vec q) =
\frac{1}{4\pi^3} \left< A^{ia}(\vec q)\, A^{ja}(-\vec q)\right>~.
\ee
The trace of $xG^{ij}_{\text{WW}}$ defines the conventional WW gluon
distribution $xG^{(1)}(x,\vec q)$ while the traceless part corresponds
to the distribution of linearly polarized gluons
$xh_\perp^{(1)}(x,\vec q)$. Both are integrated over impact parameters
since we consider the forward limit. In the non-forward case the
general decomposition of the WW GTMD involves additional independent
functions on the r.h.s.\ of eq.~(\ref{eq:WW-Gij}), see
e.g.\ ref.~\cite{Boussarie:2018zwg}.

The field in light-cone gauge is obtained from $A^+$ by a gauge
transformation,
\be \label{eq:cov--LC}
A^i(\vec x_T) = \frac{i}{g}\, U^\dagger(\vec x_T)\, \partial^i U(\vec x_T)~,
\ee
such that in this gauge $A^+(\vec x_T)=0$. At linear order in $\rho$,
$A^i(\vec q)\sim q^i\, \rho(\vec q)$ is longitudinal so that
$xG^{(1)}(x,\vec q) = xh_\perp^{(1)}(x,\vec q)$, corresponding to
maximal polarization:
\be \label{eq:xG-xh-highq}
xG^{(1)}(x,\vec q) = xh_\perp^{(1)}(x,\vec q) =
\frac{N_c^2-1}{8\pi^3 q^2} \, g^2\, G_2(\vec q, -\vec q)~.
\ee

Beyond leading order in $\rho$ (or $A^+$) the L.C.\ gauge field is no
longer purely longitudinal and one finds that $xG^{(1)}(x,\vec q) >
xh_\perp^{(1)}(x,\vec q)$. See
refs.~\cite{Metz:2011wb,Dominguez:2011br} for computations of these
distributions to all orders in $A^+$, in the Gaussian MV model of
classical color charges. Resummed WW gluon distributions for Gaussian
color charge fluctuations with a more general two-point correlator
have been derived in ref.~\cite{Dumitru:2016jku}; also see
appendix~\ref{sec:WW_Orho4}.

Here, we express the correction to
$xG^{(1)}(x,\vec q)$ and $xh_\perp^{(1)}(x,\vec q)$ at fourth order in
$A^+$ in terms of the quartic color charge correlator:
\bea
\Delta xG^{(1)}(x,\vec q) = - \Delta xh_\perp^{(1)}(x,\vec q) &=&
\frac{g^2}{16\pi^3} f^{abe} f^{cde} \int_{k,p}
\frac{1}{k^2}\frac{1}{p^2}\frac{1}{(\vec q-\vec k)^2}\frac{1}{(\vec q
  + \vec p)^2}
\left(\frac{\vec k\cdot \vec q \,\,
  \vec p\cdot \vec q}{q^2} - \vec k\cdot \vec p\right)
 \nn\\
& & \left<
\rho^a(\vec q-\vec k)\, \rho^b(\vec k)\, \rho^c(-\vec q-\vec p)\,
\rho^d(\vec p) \right>~.
\eea
The explicit expression for $f^{abe} f^{cde} \left< \rho^a(\vec q-\vec k)\,
\rho^b(\vec k)\, \rho^c(-\vec q-\vec p)\, \rho^d(\vec p) \right>$ in
terms of the proton LFwf is given in eq.~(\ref{eq:dxG_dxh_fg}) of
appendix~\ref{sec:WW_Orho4}. Hence, at this order there is a splitting
of $xG^{(1)}$ and $xh_\perp^{(1)}$ which are no longer equal.

\begin{figure}[htb]
  \centering
  \includegraphics[width=0.45\textwidth]{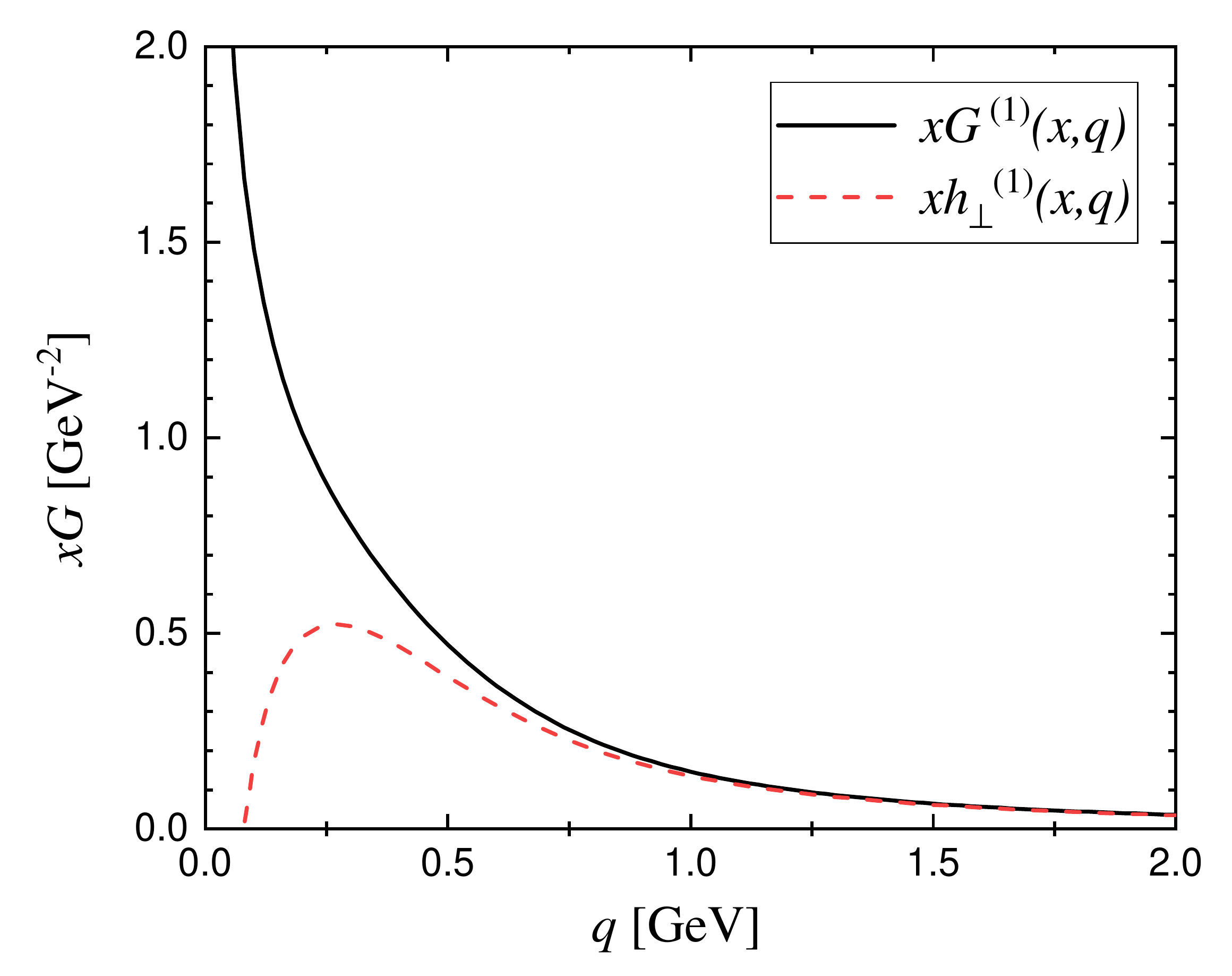}
  \includegraphics[width=0.45\textwidth]{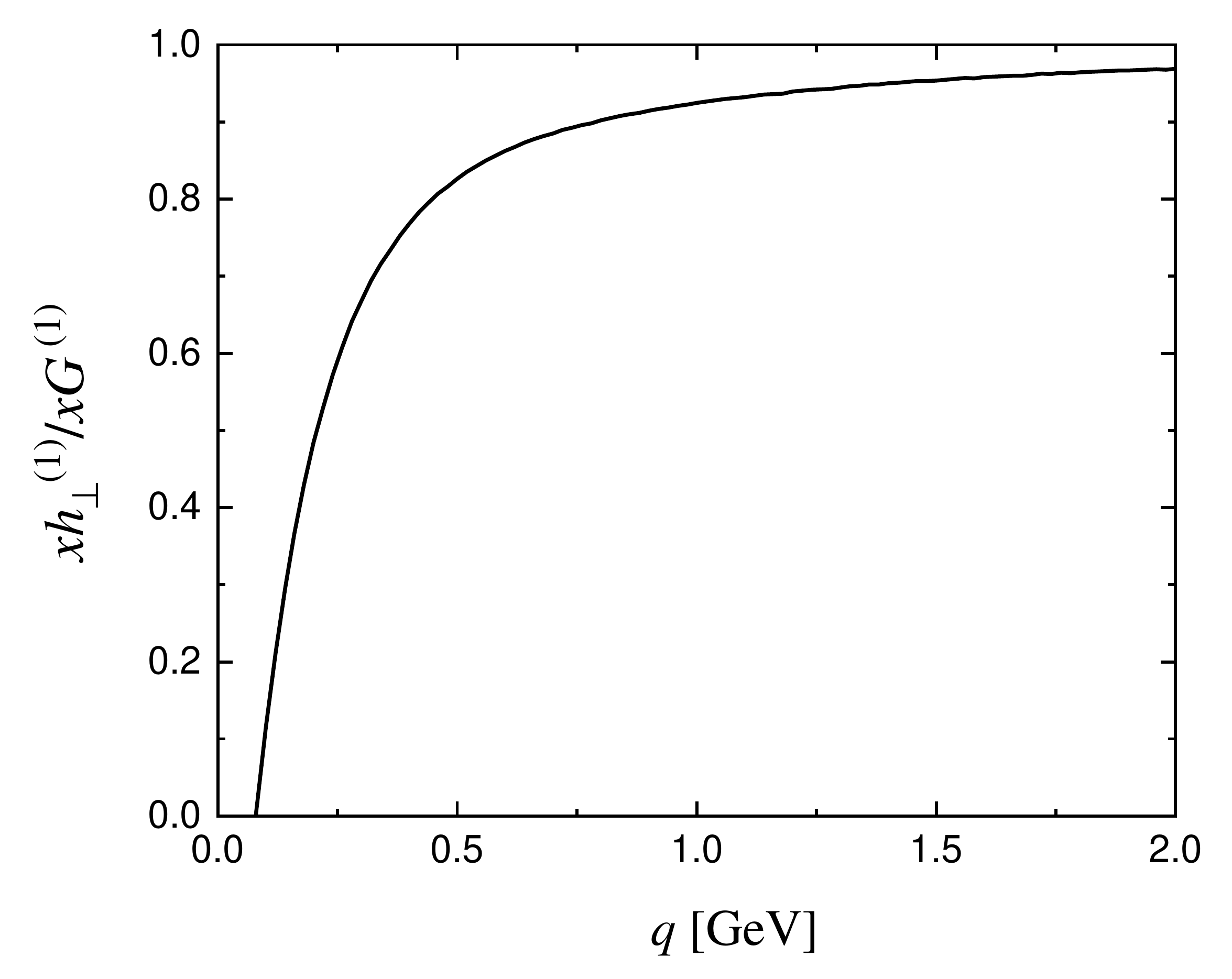}
  \caption{The conventional and linearly polarized WW gluon
    distributions in the proton (at $x\sim 0.1$) to order $(A^+)^4$.}
\label{fig:WW}
\end{figure}
Fig.~\ref{fig:WW} shows numerical results for the two WW gluon
distributions in the proton. For $q \simge 0.5$~GeV the higher twist
correction is very small and the ``polarization'' is nearly
maximal. This confirms that a measurement of $xh_\perp^{(1)}(x,q)$ at
an EIC appears promising, for example via dijet azimuthal
asymmetries~\cite{WW-dijet-smallx}. The higher twist correction
overwhelms the leading contribution below $q \sim 0.2$~GeV where a
resummation to all orders in $A^+$ would be required.  For the
Gaussian MV model of classical color charge fluctuations this has been
done in refs.~\cite{Metz:2011wb,Dominguez:2011br} (and its evolution
to small $x$ in refs.~\cite{Dumitru:2015gaa,Marquet:2016cgx}) but here
higher order correlators are independent functions and a resummation
appears difficult.

\section{Summary and Discussion} \label{sec:Summary}

\begin{figure}[htb]
  \centering
  \includegraphics[width=0.45\textwidth]{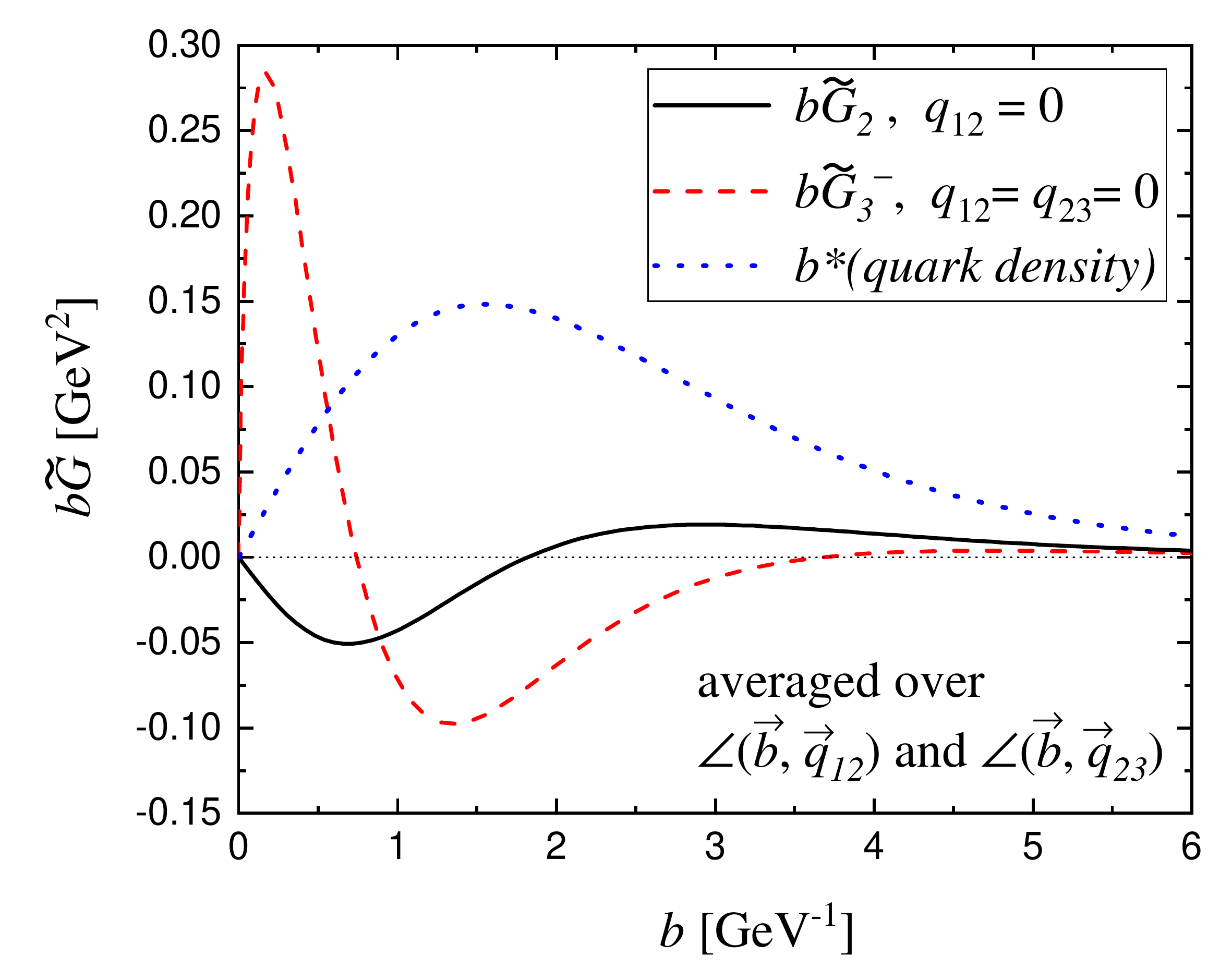}
  \caption{Quadratic and $C$-odd cubic color charge correlators, and
    the 1-body quark density, as functions of impact parameter.
}
\label{fig:G2-G3-qDensity}
\end{figure}

In this paper we have computed 2d color charge density correlations in
the proton at moderate $x\sim0.1$. The correlators of two, three and
four color charge density operators $\rho^a$ have been related
explicitly to the light-front wave function of the proton.  These
correlators exhibit interesting dependence on the relative momenta of
the probes, and on impact parameter. The two-point correlator
$G_2(\vec q_1, \vec q_2) \sim \langle\rho^a(\vec q_1)\, \rho^a(\vec
q_2)\rangle$, for example, is positive at large relative momentum
$\vec q_{12} = \vec q_1 - \vec q_2$, indicating ``attraction'' of like
charges. It turns negative (``repulsion'') at smaller relative
momentum, for central impact parameters. The correlation function
satisfies a sum rule such that at $q_{12} =0$ its integral over the
impact parameter plane vanishes: $\int \dd^2b \, \widetilde G_2(\vec
b, q_{12}=0) = 0$.  We note that $\widetilde G_2(\vec b, \vec q_{12})$
is a {\em two-body} Generalized Parton Distribution (GPD) which
depends not only on impact parameter but also on the relative
transverse momentum (or distance) of the two gluon probes\footnote{For
  the  proton wave function considered here, there is no
  dependence on $x$. We refer to ref.~\cite{Diehl:2003ny} for a review
  on GPDs.}:
\bea \label{eq:G2(b)_2-body}
\widetilde G_2(\vec b, \vec q_{12}) &=& \int_{K_T} e^{-i \vec b \cdot \vec K_T}
 \int \dd x_1 \dd  x_2
\dd x_3 \, \delta(1-x_1-x_2-x_3)
\int \frac{\dd^2 p_1 \dd^2 p_2 \dd^2 p_3}{(16\pi^3)^2}
\, \delta(\vec{p}_1+\vec{p}_2+\vec{p}_3) \nn\\
& &~~~~
\left[\psi^*(\vec p_1 +(1-x_1) \vec K_T, \vec p_2 -x_2 \vec K_T,
  \vec p_3 -x_3 \vec K_T) \right.\nn\\
  & & ~~~~\left.
  -\psi^*(\vec p_1 -\frac{\vec q_{12}-\vec K_T}{2} -x_1\vec K_T, \vec
  p_2 + \frac{\vec q_{12}+\vec K_T}{2} -x_2 \vec K_T, \vec p_3 -x_3
  \vec K_T) \right] \psi(\vec p_1, \vec p_2, \vec p_3)~.
\eea
$\psi$ denotes the amplitude of the three-quark Fock state of the
proton.  The first, one-body term is dominant for large $b$ and
$q_{12}$ while the second, two-body contribution dominates for small
$b$ and $q_{12}$.  To illustrate the importance of $n$-body
contributions to the color charge correlators, in
fig.~\ref{fig:G2-G3-qDensity} we compare $\widetilde G_2(\vec b,
q_{12}=0)$ and $\widetilde G_3^-(\vec b, q_{12}=q_{23}=0)$ to the
1-body quark density\footnote{The quark density is given by three
  times the first term in eq.~(\ref{eq:G2(b)_2-body}).} in impact
parameter space, i.e.\ to the proton ``thickness function''
$T_p(b)$. Even at vanishing relative momenta these coincide only at
rather large $b$.  The color charge correlators $\langle\rho^a(\vec
q_1)\, \rho^b(\vec q_2)\rangle$ and $\langle\rho^a(\vec q_1)\,
\rho^b(\vec q_2) \, \rho^c(\vec q_3)\rangle_{C=-}$ can be probed in
exclusive production of various charmonium states in (virtual) photon
-- proton scattering~\cite{Dumitru:2019qec,Mantysaari:2016ykx} or via
charge asymmetries in pion pair production~\cite{Hagler:2002nh}.
\\~~\\

Another main result of the paper is that color charge fluctuations in
the proton are far from Gaussian. The magnitudes of the $C$-even and
$C$-odd components of the cubic correlator $\langle \rho^a\rho^b\rho^c
\rangle / g^3$ are comparable to that of the two-point correlator
$\langle \rho^a\rho^b \rangle / g^2$. In particular, $C$-odd
correlations of cubic fluctuations near the center of the proton are
large and positive, for sufficiently small relative momenta of the
gluon probes. \\~~\\

Sub-femto-scale color charge correlations in the proton determine the
dipole scattering amplitude.  Relating them to the proton LFwf, which
could in principle be determined via ``imaging'' of the proton at a
future electron-ion collider, could help constrain and improve initial
conditions for small-$x$ evolution. In particular, our analysis
provides initial conditions which account for the above-mentioned
non-trivial structure of two- and three-point correlators as functions
of the transverse momentum ($\vec q_{12}$) or distance scale ($\vec
r$), impact parameter $\vec b$, and their relative angular
orientation. Hence, they may be useful for checking the consistency of
BK evolution with the impact parameter dependence of the dipole
$S$-matrix extracted from data at small $x$~\cite{S(b)-small-x}.

The scattering amplitude derived here also includes a non-zero $C$-odd
``Odderon'' contribution to the dipole scattering amplitude which may
be evolved to smaller $x$~\cite{Kovchegov:2003dm} to predict cross
sections for exclusive processes involving $C$-odd exchanges, or the
dipole gluon Sivers function of a transversely polarized
proton~\cite{Yao:2018vcg}. Somewhat surprisingly, perhaps, our
numerical analysis indicates that the $C$-odd amplitude for three
gluon exchange ${\cal T}_{ggg}(\vec r, \vec b)$ is much smaller in
magnitude than the $C$-even amplitude ${\cal T}_{gg}(\vec r, \vec b)$
for two gluon exchange. As already mentioned, this is not because
color charge fluctuations in the proton are nearly Gaussian. Neither
is it due to the additional power of $\alpha_s$ in ${\cal
  T}_{ggg}(\vec r, \vec b)$ which is compensated by other numerical
factors. Rather, it is mainly a consequence of the fact that this
amplitude is odd under parity. This leads to large cancellations in
the three gluon exchange diagram (for central impact parameters) when
their transverse momenta are reversed. ${\cal T}_{ggg}(\vec r, \vec
b)$ must vanish, also, for large impact parameters or large dipoles as
the density of color charge in the periphery of the proton is
low. Consequently, we expect that cross sections for semi-hard
exclusive processes involving $C$-odd three gluon exchange are small
and require high luminosity.\\~~\\

We have also computed the conventional and linearly polarized
Weizs\"acker-Williams gluon TMDs $xG^{(1)}(x,q)$ and
$xh_\perp^{(1)}(x,q)$ in the proton at moderately low $x\sim0.1$.  At
leading twist (order $(A^+)^2$) the field in light-cone gauge is
purely longitudinal and there is maximal polarization,
$xG^{(1)}(x,q)=xh_\perp^{(1)}(x,q)$. The first power correction
introduces a transverse part to $A^{ia}$ so that these gluon
distributions are no longer equal. The correction to $xG^{(1)}(x,q)$
and $xh_\perp^{(1)}(x,q)$ involves a correlator of four $A^+$ in the
proton. This is an independent function when color charge fluctuations
are not Gaussian, and we have related it explicitly to overlap
integrals of the LFwf of the proton. Numerically, we find that for
$q\simge0.5$~GeV the higher twist correction is small and
``polarization'' is close to maximal. Hence, a measurement of
$xh_\perp^{(1)}(x,q)$ at an EIC appears promising.\\

Throughout the paper we have approximated the proton in terms of its
valence quark Fock state. It will be important to include the
$\left|qqqg\right>$ Fock state, too, where the gluon is not
necessarily soft. This may affect color charge correlations which
probe high parton transverse momenta, and should improve the matching
to small-$x$ BK evolution. Work in that direction is in progress.

\section*{Acknowledgements}

We thank Y.~Hatta and L.~Motyka for useful
comments. Figs~\ref{fig:diag-rhorho}, \ref{fig:WW_O4} have
been prepared with Jaxodraw~\cite{jaxo}.

A.D.\ acknowledges support by the DOE Office of Nuclear Physics
through Grant No.\ DE-FG02-09ER41620; and from The City University of
New York through the PSC-CUNY Research grant 62098-00~50.

V.S.  acknowledges support by the DOE Office of Nuclear Physics
through Grant No. DE-SC0020081.  V.S. thanks the ExtreMe Matter Institute EMMI (GSI
Helmholtzzentrum f\"ur Schwerionenforschung, Darmstadt, Germany) for partial support and hospitality.

T.S.\ is supported by the Polish National Science Center (NCN) grants 
No.\ 2017/27/B/ST2/02755 and 2019/32/C/ST2/00202.

\appendix

\section{Simple model wave function}  \label{app:BrodskySchlumpf}

For numerical estimates we employ the ``harmonic oscillator''
model wave function of Brodsky and Schlumpf~\cite{Brodsky:1994fz},
\begin{eqnarray}
  \psi_{\rm H.O.}(x_1,\vec k_1; x_2, \vec k_2; x_3,\vec  k_3) &=& N_{\rm
    H.O.}\exp(-{\cal M}^2/2\beta^2)~.   \label{eq:pLFWF}
\end{eqnarray}
The invariant mass ${\cal M}$ of the configuration is given by
\begin{equation}
{\cal M}^2 = \sum_{i=1}^3 \frac{\vec k_i^2+m^2}{x_i}~.
\end{equation}
$\beta$ determines the typical transverse momentum of quarks in the
proton. The parameters $\beta$ and $m^2$ were determined in
ref.~\cite{Brodsky:1994fz} as $m=0.26$ GeV, $\beta=0.55$~GeV. The
normalization constant $N_{\rm H.O.}$ is obtained from the
normalization condition~(\ref{eq:Norm_psi3}).

The above simple model wave function allows us to perform analytically
parts of the evaluation of the correlators of $+$ color currents in
the proton, c.f.\ eqs.~(\ref{eq:rho2_Kt_LFwf}, \ref{eq:G3+_LFwf},
\ref{eq:rho3_Kt_fbc}, \ref{eq:rho4_full}). This simplifies the
numerical computations significantly.  Other models and
parameter sets can be found in
refs.~\cite{Frank:1995pv}.\\

\section{Color charge correlators}  \label{sec:rho_Correlators}

Following ref.~\cite{DMV} we introduce the color charge density
operators corresponding to the light cone plus component of the quark
currents
\bea
\rho^a(x_k\ll1,\vec k) &=&
g \sum_{i,j} \int\frac{\dd x_q}{x_q}
 \int \frac{\dd^2q}{16\pi^3}\,
 b^\dagger_{x_q,\vec q,i} \, b_{x_q,\vec k+\vec q,j} \, (t^a)_{ij} ~.
  \label{eq:rhoC} 
\eea
$b^\dagger_{q,i}$ and $b_{q,i}$ denote creation and annihilation
operators for quarks with plus momentum $q^+ = x_q P^+$, transverse
momentum $\vec q$, and color $i$. Note that this neglects
contributions from antiquarks and gluons which we assume are small at
$x_k\sim 0.1$.  We also neglect longitudinal momentum transfer to the
quarks and use the kinematic approximation where $x_k\sim 0.1 \ll1$.
This allows us to simplify the color charge operators as indicated
above.\\

The expectation value of a single color charge operator in the proton
is given by\footnote{$\left<\cdots\right>$ corresponds to
  $\langle K | \cdots | P\rangle$ stripped of the $\delta$-functions
  expressing conservation of transverse and plus momentum, e.g.\
  $\langle K |\rho^a(\vec q) | P\rangle = 16\pi^3 \, P^+
  \delta(P^+-K^+)\, \delta(\vec K_T + \vec q)\, \left<\rho^a(\vec
  q)\right>$, where we set $\vec P_T=0$ for the incoming
  proton.}
\bea
\left <\rho^a(-\vec K_T)\right> &=& g\, \tr t^a
\int \dd x_1\dd x_2 \dd x_3 \, \delta(1-x_1-x_2-x_3)
\int \frac{\dd^2 p_1 \dd^2 p_2 \dd^2 p_3}{(16\pi^3)^2}
\, \delta(\vec{p}_1+\vec{p}_2+\vec{p}_3)\nonumber\\
& &~ \psi^*(\vec p_1 + (1-x_1)\vec K_T,\vec p_2 -x_2\vec K_T,\vec p_3
-x_3\vec K_T) ~ \psi(\vec p_1, \vec p_2, \vec p_3) \nonumber\\
&=& g\, \tr t^a \int \dd x_1 \int \frac{\dd^2 p_1}{(2\pi)^2}\,
W^{(1)}_{K_T}(x_1,\vec p_1+\vec K_T) ~.   \label{eq:<rho>}
\eea
For brevity we omit the momentum fractions $x_1, x_2, x_3$ from the
list of arguments of $\psi$ and $\psi^*$ since we employ the eikonal
approximation. Here, $W^{(1)}_{K_T}(x_1,\vec p_1+\vec K_T)$ is the
one-body quark GTMD / Wigner distribution for momentum transfer $K_T$;
one may Fourier transform it from $\vec K_T$-space to $\vec
b$-space. Of course, $\left <\rho^a(-\vec K_T)\right>$ vanishes due to
color neutrality.\\

The correlator of two color charge density operators
is given by~\cite{DMV},
\bea
\left <\rho^a(\vec q_1) \, \rho^b(\vec q_2) \,\right>
&=& g^2\, \tr t^a t^b \int \dd x_1 \dd  x_2
\dd x_3 \, \delta(1-x_1-x_2-x_3)
\int \frac{\dd^2 p_1 \dd^2 p_2 \dd^2 p_3}{(16\pi^3)^2}
\, \delta(\vec{p}_1+\vec{p}_2+\vec{p}_3) \nn\\
& &~~~~
\left[\psi^*(\vec p_1 -\vec q_1-\vec q_2-x_1 \vec K_T, \vec p_2 -x_2 \vec K_T,
  \vec p_3 -x_3 \vec K_T) \right.\nn\\
  & & ~~~~\left.
  -\psi^*(\vec p_1 -\vec q_1 -x_1\vec K_T, \vec p_2 -\vec q_2 -x_2
  \vec K_T, \vec p_3 -x_3 \vec K_T) \right]
\psi(\vec p_1, \vec p_2, \vec p_3) \label{eq:rho2_Kt_LFwf}\\
&\equiv& \frac{1}{2}\delta^{ab}\, g^2\, G_2(\vec q_1, \vec q_2)~.
\label{eq:rho2_Kt}
\eea
$\vec K_T$ is the total momentum transfer to the proton; by
conservation of transverse momentum we have that $\vec K_T = - (\vec
q_1 + \vec q_2)$. Similarly, in all charge correlators below $\vec K_T
+ \sum_i \vec q_i =0$. Up to a conventional factor of $(-i)^2$ which we
write explicitly in the exponent of the Wilson
lines~(\ref{eq:WilsonLines}), this result coincides with the two-gluon
exchange proton impact factor given in
refs.~\cite{Bartels:2007aa,Fukugita:1978fe}.

In the limit where all $q_i$ far exceed the typical transverse
momentum of quarks in the proton, while $K_T\ll q_i$, this
correlator, as well as higher correlators introduced below, approach a
universal limit given by a one-body GPD:
\bea
G_2(\vec q_1, \vec q_2) & \to &
\int \dd x_1 \dd  x_2
\dd x_3 \, \delta(1-x_1-x_2-x_3)
\int \frac{\dd^2 p_1 \dd^2 p_2 \dd^2 p_3}{(16\pi^3)^2}
\, \delta(\vec{p}_1+\vec{p}_2+\vec{p}_3) \nn\\
& & 
 \psi^*(\vec p_1 -\vec q_1 - \vec q_2 -x_1 \vec K_T, \vec p_2 -x_2 \vec K_T,
  \vec p_3 -x_3 \vec K_T) ~
  \psi(\vec p_1, \vec p_2, \vec p_3) \nn\\
  &=& \int \dd x_1 \int \frac{\dd^2 p_1}{(2\pi)^2}\,
  W^{(1)}_{K_T}(x_1,\vec p_1 + \vec K_T)~,~~~~~~~~~~~~~~~~~~~~~~~~~
  (K_T \ll q_1, q_2)~.
  \label{eq:F(Kt)}
\eea
The term ``one-body GPD'' refers to the fact that both color charge
operators act on one and the same quark and one may integrate out the
spectator quarks. On the other hand, when the probes share a
large momentum transfer $\vec K_T$ the dominant contribution
is due to the diagram where the two gluons attach to
different quarks in the proton, i.e.\ to the two-body representation
of $\rho^a(\vec q_1)\, \rho^b(\vec q_2)$ which gives the second term
in eq.~(\ref{eq:rho2_Kt_LFwf})~\cite{Dumitru:2019qec}:
\bea
G_2(\vec q_1, \vec q_2) & \to &
-\int \dd x_1 \dd  x_2
\dd x_3 \, \delta(1-x_1-x_2-x_3)
\int \frac{\dd^2 p_1 \dd^2 p_2 \dd^2 p_3}{(16\pi^3)^2}
\, \delta(\vec{p}_1+\vec{p}_2+\vec{p}_3) \nn\\
& &
\psi^*(\vec p_1 -\vec q_1 -x_1\vec K_T, \vec p_2 -\vec q_2 -x_2
  \vec K_T, \vec p_3 -x_3 \vec K_T)\,
\psi(\vec p_1, \vec p_2, \vec p_3)\nn\\
  &=& \int \dd x_1 \int \frac{\dd^2 p_1}{(2\pi)^2}\int \dd x_2 \int
\frac{\dd^2 p_2}{(2\pi)^2}\,
  W^{(2)}_{K_T}(x_1,\vec p_1 - \vec q_1, x_2, \vec p_2 - \vec q_2)
  ~,~~~~~~~~~~~~~~~~~~(\vec q_1, \vec q_2 \sim -\vec K_T/2)~.
\label{eq:rho2_large-Kt}
\eea
This involves a two-body GTMD or Wigner distribution. The $n$-body
diagrams are important for exclusive photo-production of charmonium at
large $-t$~\cite{Dumitru:2019qec}.
\\

We now proceed with cubic and quartic color charge correlators. The
fact that $\langle \rho^a(\vec q_1) \, \rho^b(\vec q_2) \, \rho^c(\vec
q_3) \,\rangle$ is not zero shows that color charge fluctuations are
not Gaussian.
The $C$-odd part of the cubic correlator is given by~\cite{DMV}
\bea
\left< \rho^a(\vec q_1) \, \rho^b(\vec q_2) \, \rho^c(\vec
q_3) \,\right>_{C=-}
&=& \,\frac{1}{4}\,d^{abc}\, g^3\,
\int \dd x_1 \dd x_2 \dd x_3 \,
\delta(1-x_1-x_2-x_3) \int \frac{\dd^2 p_1 \dd^2 p_2
  \dd^2 p_3}{(16\pi^3)^2} \, \delta(\vec{p}_1+\vec{p}_2+\vec{p}_3)
\nn\\
& &
\left[ 
\psi^*(\vec{p}_1-\vec q_1-\vec q_2-\vec q_3  -x_1\vec K_\perp ,\vec{p}_2-x_2\vec
K_\perp, \vec{p}_3-x_3\vec K_\perp) \right. \nonumber\\
& & -
  \psi^*(\vec{p}_1-\vec q_1 -x_1\vec K_\perp ,\vec{p}_2-\vec q_2-\vec q_3
  - x_2\vec K_\perp, \vec{p}_3-x_3\vec K_\perp) \nonumber\\
  & &
  -
 \psi^*(\vec{p}_1-\vec q_1-\vec q_3 -x_1\vec K_\perp ,\vec{p}_2-\vec q_2
  - x_2\vec K_\perp, \vec{p}_3-x_3\vec K_\perp) \nonumber\\ & & -
 \psi^*(\vec{p}_1-\vec q_1-\vec q_2 -x_1 \vec K_\perp
  ,\vec{p}_2-\vec q_3- x_2\vec K_\perp,
  \vec{p}_3-x_3\vec K_\perp) \nonumber\\ & & +2\, 
\psi^*(\vec{p}_1-\vec
  q_1 -x_1\vec K_\perp ,\vec{p}_2-\vec q_2 -x_2 \vec K_\perp,
  \vec{p}_3-\vec q_3 - x_3\vec K_\perp ) \nn\\
  & & \Bigr]\, \psi(\vec p_1, \vec p_2 ,\vec p_3) \label{eq:rho^3correl}\\
&\equiv&
\frac{1}{4}d^{abc}\, g^3\, G_3^-(\vec q_1, \vec q_2, \vec q_3)~.
\label{eq:rho3_Kt}
\eea
Again, this expression agrees with the $C$-odd three-gluon
exchange proton impact factor $E_{3;0}$ by Bartels and
Motyka~\cite{Bartels:2007aa} (also see refs.~\cite{Czyzewski:1996bv})
up to a conventional factor of $(-i)^3$.

$G_3^-$ can be expressed in terms of 2-gluon exchange correlators
$G_2$, where two of the three gluons are ``paired up'',
plus a genuine 3-body contribution which enforces the Ward identity
(vanishing of $G_3^-$) when either one $\vec q_i \to 0$:
\bea
G_3^-(\vec q_1, \vec q_2, \vec q_3) &=&
G_2(\vec q_1+ \vec q_2, \vec q_3) 
+G_2(\vec q_1+ \vec q_3, \vec q_2) 
+G_2(\vec q_2+ \vec q_3, \vec q_1) \nn\\
& & -2\int \dd x_1 \dd x_2 \dd x_3 \,
\delta(1-x_1-x_2-x_3) \int \frac{\dd^2 p_1 \dd^2 p_2
  \dd^2 p_3}{(16\pi^3)^2} \, \delta(\vec{p}_1+\vec{p}_2+\vec{p}_3)\nn\\
& & ~~~~~~~~~~\left[
\psi^*(\vec{p}_1-\vec q_1-\vec q_2-\vec q_3  -x_1\vec K_\perp ,\vec{p}_2-x_2\vec
K_\perp, \vec{p}_3-x_3\vec K_\perp) \right. \nonumber\\
& & ~~~~~~~~~ \left. -\psi^*(\vec{p}_1-\vec
  q_1 -x_1\vec K_\perp ,\vec{p}_2-\vec q_2 -x_2 \vec K_\perp,
  \vec{p}_3-\vec q_3 - x_3\vec K_\perp ) \right]
  \, \psi(\vec p_1, \vec p_2 ,\vec p_3)~.
\eea
\\

For completeness we also give the $C$-even (or negative signature)
part of the cubic correlator although it is not needed for the dipole
scattering amplitude:
\bea
\left< \rho^a(\vec q_1) \, \rho^b(\vec q_2) \, \rho^c(\vec
q_3) \,\right>_{C=+}
&=& \,\frac{i}{4}\,f^{abc}\,g^3\,
\int \dd x_1 \dd x_2 \dd x_3 \,
\delta(1-x_1-x_2-x_3) \int \frac{\dd^2 p_1 \dd^2 p_2
  \dd^2 p_3}{(16\pi^3)^2} \, \delta(\vec{p}_1+\vec{p}_2+\vec{p}_3)
\nn\\
& &
\left[ 
\psi^*(\vec{p}_1-\vec q_1-\vec q_2-\vec q_3-x_1\vec K_\perp ,\vec{p}_2-x_2\vec
K_\perp, \vec{p}_3-x_3\vec K_\perp) \right. \nonumber\\
& & -
  \psi^*(\vec{p}_1-\vec q_2-\vec q_3-x_1\vec K_\perp ,\vec{p}_2-\vec q_1
  - x_2\vec K_\perp, \vec{p}_3-x_3\vec K_\perp) \nonumber\\
  & &
  +
 \psi^*(\vec{p}_1 -\vec q_1-\vec q_3-x_1\vec K_\perp ,\vec{p}_2-\vec q_2
 - x_2\vec K_\perp, \vec{p}_3-x_3\vec K_\perp) \nonumber\\
 & & -
 \psi^*(\vec{p}_1-\vec q_1-\vec q_2-x_1 \vec K_\perp
  ,\vec{p}_2-\vec q_3- x_2\vec K_\perp,
  \vec{p}_3-x_3\vec K_\perp)\nn\\
  & & \Bigr]\, \psi(\vec p_1, \vec p_2 ,\vec p_3) \label{eq:G3+_LFwf}\\
&\equiv&
\frac{i}{4}f^{abc}\, g^3\, G_3^+(\vec q_1, \vec q_2, \vec q_3)~.
\label{eq:rho3_Kt_fbc}
\eea
$G_3^+$ can be fully decomposed into 2-gluon exchanges, similar to
Reggeized gluon exchanges at
small-$x$~\cite{Ewerz:2001fb,Bartels:2007aa}:
\be \label{eq:G3+_Reggeized}
G_3^+(\vec q_1, \vec q_2, \vec q_3) = G_2(\vec q_1+ \vec q_2, \vec q_3)
- G_2(\vec q_1+ \vec q_3, \vec q_2) +
G_2(\vec q_1, \vec q_2+ \vec q_3)~.
\ee
This vanishes when the transverse momentum of the first or last gluon
($\vec q_1$ resp.\ $\vec q_3$) is taken to zero but not for $\vec
q_2\to 0$~\cite{Ewerz:2001fb}.
\\

Lastly, the correlator of four color charge operators is given by
\begin{eqnarray}
& & \left< \rho^a(\vec q_1)\,\rho^b(\vec q_2)\, \rho^c(\vec
  q_3)\, \rho^d(\vec q_4)\,  \right> = g^4 \nn\\
  & &
  \int \dd x_1\, \dd x_2\, \dd x_3\, \delta(1- x_1-x_2-x_3)
  \int \frac{\dd^2 p_1\,\dd^2 p_2\,\dd^2 p_3}{(16\pi^3)^2}\,
  \delta(\vec p_1 + \vec p_2 + \vec p_3) \,
  \psi(\vec p_1,\vec p_2,\vec p_3) \nn\\
& & \left\{
  \tr t^a t^b t^c t^d ~ \psi^*(\vec p_1 - \vec q_1 - \vec q_2 - \vec
  q_3 - \vec q_4 - x_1\vec K_T, \vec
  p_2-x_2\vec K_T, \vec p_3-x_3\vec K_T) \right.\nn\\
& &
+ \left( \tr t^a t^b \, \tr t^c t^d - \tr t^a t^b t^c t^d\right)  ~
\psi^*(\vec p_1-\vec q_1-\vec q_2-x_1\vec K_T, \vec p_2-\vec q_3-\vec
       q_4-x_2\vec K_T, \vec p_3-x_3\vec K_T)\nn\\
       & &
       + \left( \tr t^a t^c \, \tr t^b t^d - \tr t^a t^c t^b t^d\right)  ~
\psi^*(\vec p_1-\vec q_1-\vec q_3-x_1\vec K_T, \vec p_2-\vec q_2-\vec
       q_4-x_2\vec K_T, \vec p_3-x_3\vec K_T)\nn\\
       & &
       + \left( \tr t^a t^d \, \tr t^b t^c - \tr t^a t^d t^b t^c\right)  ~
\psi^*(\vec p_1-\vec q_1-\vec q_4-x_1\vec K_T, \vec p_2-\vec q_2-\vec
       q_3-x_2\vec K_T, \vec p_3-x_3\vec K_T)\nn\\
       & &
       - \tr t^a t^b t^c t^d ~ \psi^*(\vec p_1-\vec q_1-\vec q_2-\vec
       q_3-x_1\vec K_T, \vec p_2-\vec q_4-x_2\vec K_T, \vec
       p_3-x_3\vec K_T) \nn \\
       & &
       - \tr t^a t^b t^c t^d ~ \psi^*(\vec p_1-\vec q_1-x_1\vec K_T,
       \vec p_2-\vec q_2-\vec q_3-\vec q_4-x_2\vec K_T, \vec
       p_3-x_3\vec K_T) \nn \\
       & &
       - \tr t^a t^b t^d t^c ~ \psi^*(\vec p_1-\vec q_1-\vec q_2-\vec
       q_4-x_1\vec K_T, \vec p_2-\vec q_3-x_2\vec K_T, \vec
       p_3-x_3\vec K_T) \nn \\
       & &
       - \tr t^a t^c t^d t^b ~ \psi^*(\vec p_1-\vec q_1-\vec q_3-\vec
       q_4-x_1\vec K_T, \vec p_2-\vec q_2-x_2\vec K_T, \vec
       p_3-x_3\vec K_T) \nn \\
       & &
       + \left(\tr t^a t^b t^c t^d+\tr t^a t^b t^d t^c- \tr t^a t^b \,
       \tr t^c t^d\right) ~ \psi^*(\vec p_1-\vec q_1-\vec q_2-x_1\vec
       K_T, \vec p_2-\vec q_3-x_2\vec K_T, \vec p_3-\vec q_4-x_3\vec K_T) \nn\\
       & &
       + \left(\tr t^a t^c t^b t^d+\tr t^a t^c t^d t^b- \tr t^a t^c \,
       \tr t^b t^d\right) ~ \psi^*(\vec p_1-\vec q_1-\vec q_3-x_1\vec
       K_T, \vec p_2-\vec q_2-x_2\vec K_T, \vec p_3-\vec q_4-x_3\vec K_T) \nn\\
       & &
       + \left(\tr t^a t^d t^b t^c+\tr t^a t^d t^c t^b- \tr t^a t^d \,
       \tr t^b t^c\right) ~ \psi^*(\vec p_1-\vec q_1-\vec q_4-x_1\vec
       K_T, \vec p_2-\vec q_2-x_2\vec K_T, \vec p_3-\vec q_3-x_3\vec K_T) \nn\\
       & &
       + \left(\tr t^a t^b t^c t^d+\tr t^a t^d t^b t^c- \tr t^a t^d \,
       \tr t^b t^c\right) ~ \psi^*(\vec p_1-\vec q_1-x_1\vec K_T, \vec
       p_2-\vec q_2-\vec q_3-x_2\vec K_T, \vec p_3-\vec q_4-x_3\vec K_T) \nn\\
       & &
       + \left(\tr t^a t^b t^c t^d+\tr t^a t^c t^d t^b- \tr t^a t^b \,
       \tr t^c t^d\right) ~ \psi^*(\vec p_1-\vec q_1-x_1\vec K_T, \vec
       p_2-\vec q_2-x_2\vec K_T, \vec p_3-\vec q_3-\vec q_4-x_3\vec K_T) \nn\\
       & &
       + \left(\tr t^a t^b t^d t^c+\tr t^a t^c t^b t^d- \tr t^a t^c \,
       \tr t^b t^d\right) ~ \psi^*(\vec p_1-\vec q_1-x_1\vec K_T, \vec
       p_2-\vec q_2-\vec q_4-x_2\vec K_T, \vec p_3-\vec q_3-x_3\vec K_T)
       \Bigr\}~,    \label{eq:rho4_full}
\end{eqnarray}
where $\vec K_T \equiv - (\vec q_1+\vec q_2+\vec q_3+\vec q_4)$. Note
that it is not equal to a sum over all permutations of pairwise
contractions, confirming that color charge fluctuations are not
Gaussian.

We can decompose this correlator into $C$-even and odd parts. Charge
conjugation transforms $t^a \to - t^{aT}$ so that $\tr t^a t^b t^c t^d
\to \tr t^d t^c t^b t^a = \tr t^b t^a t^d t^c$ which corresponds to
the permutations $a\leftrightarrow b, c\leftrightarrow d$. Hence,
using $\tr t^a t^b t^c t^d = (1/12)\delta^{ab}\delta^{cd} + (1/8)
(d^{abe} + i f^{abe}) (d^{cde} + i f^{cde})$ we see that the $C$-even
pieces of $\tr t^a t^b t^c t^d$ correspond to the color structures
$\delta^{ab}\delta^{cd}$, $d^{abe}d^{cde}$, and $f^{abe}f^{cde}$;
while the $C$-odd pieces correspond to $id^{abe}f^{cde}$.

Therefore, the $C$-even parts of $\langle\rho^4\rangle$ are:
\begin{eqnarray}
& & \left< \rho^a(\vec q_1)\,\rho^b(\vec q_2)\, \rho^c(\vec
  q_3)\, \rho^d(\vec q_4)\,  \right>_{ff} = -\frac{1}{8}\,g^4 \nn\\
  & &
  \int \dd x_1\, \dd x_2\, \dd x_3\, \delta(1- x_1-x_2-x_3)
  \int \frac{\dd^2 p_1\,\dd^2 p_2\,\dd^2 p_3}{(16\pi^3)^2}\,
  \delta(\vec p_1 + \vec p_2 + \vec p_3) \,
  \psi(\vec p_1,\vec p_2,\vec p_3) \nn\\
& & \left\{
  f^{abe}f^{cde} ~ \psi^*(\vec p_1 - \vec q_1 - \vec q_2 - \vec
  q_3 - \vec q_4 - x_1\vec K_T, \vec
  p_2-x_2\vec K_T, \vec p_3-x_3\vec K_T) \right.\nn\\
& &
 -   f^{abe}f^{cde} ~
\psi^*(\vec p_1-\vec q_1-\vec q_2-x_1\vec K_T, \vec p_2-\vec q_3-\vec
       q_4-x_2\vec K_T, \vec p_3-x_3\vec K_T)\nn\\
       & &
-   f^{ace}f^{bde} ~
\psi^*(\vec p_1-\vec q_1-\vec q_3-x_1\vec K_T, \vec p_2-\vec q_2-\vec
       q_4-x_2\vec K_T, \vec p_3-x_3\vec K_T)\nn\\
       & &
-   f^{ade}f^{bce}  ~
\psi^*(\vec p_1-\vec q_1-\vec q_4-x_1\vec K_T, \vec p_2-\vec q_2-\vec
       q_3-x_2\vec K_T, \vec p_3-x_3\vec K_T)\nn\\
       & &
-   f^{abe}f^{cde} ~ \psi^*(\vec p_1-\vec q_1-\vec q_2-\vec
       q_3-x_1\vec K_T, \vec p_2-\vec q_4-x_2\vec K_T, \vec
       p_3-x_3\vec K_T) \nn \\
       & &
-   f^{abe}f^{cde} ~ \psi^*(\vec p_1-\vec q_1-x_1\vec K_T,
       \vec p_2-\vec q_2-\vec q_3-\vec q_4-x_2\vec K_T, \vec
       p_3-x_3\vec K_T) \nn \\
       & &
-   f^{abe}f^{dce} ~ \psi^*(\vec p_1-\vec q_1-\vec q_2-\vec
       q_4-x_1\vec K_T, \vec p_2-\vec q_3-x_2\vec K_T, \vec
       p_3-x_3\vec K_T) \nn \\
       & &
-   f^{ace}f^{dbe} ~ \psi^*(\vec p_1-\vec q_1-\vec q_3-\vec
       q_4-x_1\vec K_T, \vec p_2-\vec q_2-x_2\vec K_T, \vec
       p_3-x_3\vec K_T) \nn \\
       \Bigr\}~,    \label{eq:rho4_C+_ff}
\end{eqnarray}
\begin{eqnarray}
& & \left< \rho^a(\vec q_1)\,\rho^b(\vec q_2)\, \rho^c(\vec
  q_3)\, \rho^d(\vec q_4)\,  \right>_{dd} = \frac{1}{8}\,g^4 \nn\\
  & &
  \int \dd x_1\, \dd x_2\, \dd x_3\, \delta(1- x_1-x_2-x_3)
  \int \frac{\dd^2 p_1\,\dd^2 p_2\,\dd^2 p_3}{(16\pi^3)^2}\,
  \delta(\vec p_1 + \vec p_2 + \vec p_3) \,
  \psi(\vec p_1,\vec p_2,\vec p_3) \nn\\
& & \left\{
  d^{abe}d^{cde} ~ \psi^*(\vec p_1 - \vec q_1 - \vec q_2 - \vec
  q_3 - \vec q_4 - x_1\vec K_T, \vec
  p_2-x_2\vec K_T, \vec p_3-x_3\vec K_T) \right.\nn\\
& &
 -   d^{abe}d^{cde} ~
\psi^*(\vec p_1-\vec q_1-\vec q_2-x_1\vec K_T, \vec p_2-\vec q_3-\vec
       q_4-x_2\vec K_T, \vec p_3-x_3\vec K_T)\nn\\
       & &
-   d^{ace}d^{bde} ~
\psi^*(\vec p_1-\vec q_1-\vec q_3-x_1\vec K_T, \vec p_2-\vec q_2-\vec
       q_4-x_2\vec K_T, \vec p_3-x_3\vec K_T)\nn\\
       & &
-   d^{ade}d^{bce}  ~
\psi^*(\vec p_1-\vec q_1-\vec q_4-x_1\vec K_T, \vec p_2-\vec q_2-\vec
       q_3-x_2\vec K_T, \vec p_3-x_3\vec K_T)\nn\\
       & &
-   d^{abe}d^{cde} ~ \psi^*(\vec p_1-\vec q_1-\vec q_2-\vec
       q_3-x_1\vec K_T, \vec p_2-\vec q_4-x_2\vec K_T, \vec
       p_3-x_3\vec K_T) \nn \\
       & &
-   d^{abe}d^{cde} ~ \psi^*(\vec p_1-\vec q_1-x_1\vec K_T,
       \vec p_2-\vec q_2-\vec q_3-\vec q_4-x_2\vec K_T, \vec
       p_3-x_3\vec K_T) \nn \\
       & &
-   d^{abe}d^{dce} ~ \psi^*(\vec p_1-\vec q_1-\vec q_2-\vec
       q_4-x_1\vec K_T, \vec p_2-\vec q_3-x_2\vec K_T, \vec
       p_3-x_3\vec K_T) \nn \\
       & &
-   d^{ace}d^{bde} ~ \psi^*(\vec p_1-\vec q_1-\vec q_3-\vec
       q_4-x_1\vec K_T, \vec p_2-\vec q_2-x_2\vec K_T, \vec
       p_3-x_3\vec K_T) \nn \\
       & &
+ 2d^{abe}d^{cde} ~ \psi^*(\vec p_1-\vec q_1-\vec q_2-x_1\vec
       K_T, \vec p_2-\vec q_3-x_2\vec K_T, \vec p_3-\vec q_4-x_3\vec K_T) \nn\\
       & &
+ 2d^{ace}d^{bde} ~ \psi^*(\vec p_1-\vec q_1-\vec q_3-x_1\vec
       K_T, \vec p_2-\vec q_2-x_2\vec K_T, \vec p_3-\vec q_4-x_3\vec K_T) \nn\\
       & &
+ 2d^{ade}d^{bce} ~ \psi^*(\vec p_1-\vec q_1-\vec q_4-x_1\vec
       K_T, \vec p_2-\vec q_2-x_2\vec K_T, \vec p_3-\vec q_3-x_3\vec K_T) \nn\\
       & &
+ 2d^{bce}d^{ade}~ \psi^*(\vec p_1-\vec q_1-x_1\vec K_T, \vec
       p_2-\vec q_2-\vec q_3-x_2\vec K_T, \vec p_3-\vec q_4-x_3\vec K_T) \nn\\
       & &
+ 2d^{abe} d^{cde}~ \psi^*(\vec p_1-\vec q_1-x_1\vec K_T, \vec
       p_2-\vec q_2-x_2\vec K_T, \vec p_3-\vec q_3-\vec q_4-x_3\vec K_T) \nn\\
       & &
+ 2d^{ace} d^{bde}~ \psi^*(\vec p_1-\vec q_1-x_1\vec K_T, \vec
       p_2-\vec q_2-\vec q_4-x_2\vec K_T, \vec p_3-\vec q_3-x_3\vec K_T)
       \Bigr\}~,    \label{eq:rho4_C+_dd}
\end{eqnarray}
and
\begin{eqnarray}
& & \left< \rho^a(\vec q_1)\,\rho^b(\vec q_2)\, \rho^c(\vec
  q_3)\, \rho^d(\vec q_4)\,  \right>_{\delta\delta} = \frac{1}{12}\, g^4 \nn\\
  & &
  \int \dd x_1\, \dd x_2\, \dd x_3\, \delta(1- x_1-x_2-x_3)
  \int \frac{\dd^2 p_1\,\dd^2 p_2\,\dd^2 p_3}{(16\pi^3)^2}\,
  \delta(\vec p_1 + \vec p_2 + \vec p_3) \,
  \psi(\vec p_1,\vec p_2,\vec p_3) \nn\\
& & \left\{
  \delta^{ab}\delta^{cd} ~ \psi^*(\vec p_1 - \vec q_1 - \vec q_2 - \vec
  q_3 - \vec q_4 - x_1\vec K_T, \vec
  p_2-x_2\vec K_T, \vec p_3-x_3\vec K_T) \right.\nn\\
& &
 +   2\delta^{ab}\delta^{cd} ~
\psi^*(\vec p_1-\vec q_1-\vec q_2-x_1\vec K_T, \vec p_2-\vec q_3-\vec
       q_4-x_2\vec K_T, \vec p_3-x_3\vec K_T)\nn\\
       & &
+ 2\delta^{ac}\delta^{bd} ~
\psi^*(\vec p_1-\vec q_1-\vec q_3-x_1\vec K_T, \vec p_2-\vec q_2-\vec
       q_4-x_2\vec K_T, \vec p_3-x_3\vec K_T)\nn\\
       & &
+ 2\delta^{ad}\delta^{bc}  ~
\psi^*(\vec p_1-\vec q_1-\vec q_4-x_1\vec K_T, \vec p_2-\vec q_2-\vec
       q_3-x_2\vec K_T, \vec p_3-x_3\vec K_T)\nn\\
       & &
- \delta^{ab}\delta^{cd} ~ \psi^*(\vec p_1-\vec q_1-\vec q_2-\vec
       q_3-x_1\vec K_T, \vec p_2-\vec q_4-x_2\vec K_T, \vec
       p_3-x_3\vec K_T) \nn \\
       & &
- \delta^{ab}\delta^{cd} ~ \psi^*(\vec p_1-\vec q_1-x_1\vec K_T,
       \vec p_2-\vec q_2-\vec q_3-\vec q_4-x_2\vec K_T, \vec
       p_3-x_3\vec K_T) \nn \\
       & &
- \delta^{ab}\delta^{cd} ~ \psi^*(\vec p_1-\vec q_1-\vec q_2-\vec
       q_4-x_1\vec K_T, \vec p_2-\vec q_3-x_2\vec K_T, \vec
       p_3-x_3\vec K_T) \nn \\
       & &
- \delta^{ac}\delta^{db} ~ \psi^*(\vec p_1-\vec q_1-\vec q_3-\vec
       q_4-x_1\vec K_T, \vec p_2-\vec q_2-x_2\vec K_T, \vec
       p_3-x_3\vec K_T) \nn \\
       & &
- \delta^{ab}\delta^{cd} ~ \psi^*(\vec p_1-\vec q_1-\vec q_2-x_1\vec
       K_T, \vec p_2-\vec q_3-x_2\vec K_T, \vec p_3-\vec q_4-x_3\vec K_T) \nn\\
       & &
- \delta^{ac}\delta^{bd} ~ \psi^*(\vec p_1-\vec q_1-\vec q_3-x_1\vec
       K_T, \vec p_2-\vec q_2-x_2\vec K_T, \vec p_3-\vec q_4-x_3\vec K_T) \nn\\
       & &
- \delta^{ad}\delta^{bc} ~ \psi^*(\vec p_1-\vec q_1-\vec q_4-x_1\vec
       K_T, \vec p_2-\vec q_2-x_2\vec K_T, \vec p_3-\vec q_3-x_3\vec K_T) \nn\\
       & &
- \delta^{ad}\delta^{bc}~ \psi^*(\vec p_1-\vec q_1-x_1\vec K_T, \vec
       p_2-\vec q_2-\vec q_3-x_2\vec K_T, \vec p_3-\vec q_4-x_3\vec  K_T) \nn\\
       & &
- \delta^{ab} \delta^{cd}~ \psi^*(\vec p_1-\vec q_1-x_1\vec K_T, \vec
       p_2-\vec q_2-x_2\vec K_T, \vec p_3-\vec q_3-\vec q_4-x_3\vec K_T) \nn\\
       & &
- \delta^{ac} \delta^{bd}~ \psi^*(\vec p_1-\vec q_1-x_1\vec K_T, \vec
       p_2-\vec q_2-\vec q_4-x_2\vec K_T, \vec p_3-\vec q_3-x_3\vec K_T)
       \Bigr\}~.    \label{eq:rho4_C+_deltadelta}
\end{eqnarray}

The $C$-odd part of $\langle\rho^4\rangle$ is
\begin{eqnarray}
& & \left< \rho^a(\vec q_1)\,\rho^b(\vec q_2)\, \rho^c(\vec
  q_3)\, \rho^d(\vec q_4)\,  \right>_{fd} = \frac{i}{8}\, g^4 \nn\\
  & &
  \int \dd x_1\, \dd x_2\, \dd x_3\, \delta(1- x_1-x_2-x_3)
  \int \frac{\dd^2 p_1\,\dd^2 p_2\,\dd^2 p_3}{(16\pi^3)^2}\,
  \delta(\vec p_1 + \vec p_2 + \vec p_3) \,
  \psi(\vec p_1,\vec p_2,\vec p_3) \nn\\
& & \left\{
  \left(d^{abe}f^{cde}+f^{abe}d^{cde}\right) ~ \psi^*(\vec p_1 - \vec
  q_1 - \vec q_2 - \vec q_3 - \vec q_4 - x_1\vec K_T, \vec
  p_2-x_2\vec K_T, \vec p_3-x_3\vec K_T) \right.\nn\\
& &
- \left(d^{abe}f^{cde}+f^{abe}d^{cde}\right) ~
\psi^*(\vec p_1-\vec q_1-\vec q_2-x_1\vec K_T, \vec p_2-\vec q_3-\vec
       q_4-x_2\vec K_T, \vec p_3-x_3\vec K_T)\nn\\
       & &
- \left(d^{ace}f^{bde}+f^{ace}d^{bde}\right) ~
\psi^*(\vec p_1-\vec q_1-\vec q_3-x_1\vec K_T, \vec p_2-\vec q_2-\vec
       q_4-x_2\vec K_T, \vec p_3-x_3\vec K_T)\nn\\
       & &
- \left(d^{ade}f^{bce}+f^{ade}d^{bce}\right) ~
\psi^*(\vec p_1-\vec q_1-\vec q_4-x_1\vec K_T, \vec p_2-\vec q_2-\vec
       q_3-x_2\vec K_T, \vec p_3-x_3\vec K_T)\nn\\
       & &
- \left(d^{abe}f^{cde}+f^{abe}d^{cde}\right) ~ \psi^*(\vec p_1-\vec
q_1-\vec q_2-\vec q_3-x_1\vec K_T, \vec p_2-\vec q_4-x_2\vec K_T, \vec
       p_3-x_3\vec K_T) \nn \\
       & &
- \left(d^{abe}f^{cde}+f^{abe}d^{cde}\right) ~ \psi^*(\vec p_1-\vec q_1-x_1\vec K_T,
       \vec p_2-\vec q_2-\vec q_3-\vec q_4-x_2\vec K_T, \vec
       p_3-x_3\vec K_T) \nn \\
       & &
- \left(d^{abe}f^{dce}+f^{abe}d^{dce}\right) ~ \psi^*(\vec p_1-\vec
q_1-\vec q_2-\vec q_4-x_1\vec K_T, \vec p_2-\vec q_3-x_2\vec K_T, \vec
       p_3-x_3\vec K_T) \nn \\
       & &
- \left(d^{ace}f^{dbe}+f^{ace}d^{dbe}\right) ~ \psi^*(\vec p_1-\vec
q_1-\vec q_3-\vec q_4-x_1\vec K_T, \vec p_2-\vec q_2-x_2\vec K_T, \vec
       p_3-x_3\vec K_T) \nn \\
       & &
+ 2 f^{abe}d^{cde} ~ \psi^*(\vec p_1-\vec q_1-\vec q_2-x_1\vec
       K_T, \vec p_2-\vec q_3-x_2\vec K_T, \vec p_3-\vec q_4-x_3\vec K_T) \nn\\
       & &
+ 2 f^{ace}d^{bde} ~ \psi^*(\vec p_1-\vec q_1-\vec q_3-x_1\vec
       K_T, \vec p_2-\vec q_2-x_2\vec K_T, \vec p_3-\vec q_4-x_3\vec K_T) \nn\\
       & &
+ 2 f^{ade}d^{bce} ~ \psi^*(\vec p_1-\vec q_1-\vec q_4-x_1\vec
       K_T, \vec p_2-\vec q_2-x_2\vec K_T, \vec p_3-\vec q_3-x_3\vec K_T) \nn\\
       & &
+ 2 f^{bce}d^{ade}~ \psi^*(\vec p_1-\vec q_1-x_1\vec K_T, \vec
       p_2-\vec q_2-\vec q_3-x_2\vec K_T, \vec p_3-\vec q_4-x_3\vec K_T)\nn\\
       & &
+ 2f^{cde} d^{abe} ~ \psi^*(\vec p_1-\vec q_1-x_1\vec K_T, \vec
       p_2-\vec q_2-x_2\vec K_T, \vec p_3-\vec q_3-\vec q_4-x_3\vec K_T) \nn\\
       & &
+ 2f^{bde} d^{ace} ~ \psi^*(\vec p_1-\vec q_1-x_1\vec K_T, \vec
       p_2-\vec q_2-\vec q_4-x_2\vec K_T, \vec p_3-\vec q_3-x_3\vec K_T)
       \Bigr\}~.    \label{eq:rho4_C-_df}
\end{eqnarray}
Using SU(3) identities\footnote{See ref.~\cite{macfarlane1968}, in particular eq.~(2.22).}, we verified that
eqs.~(\ref{eq:rho4_C+_ff} - \ref{eq:rho4_C-_df}) agree with the
expressions in section~4.4 of ref.~\cite{Bartels:2007aa}.

\section{Weizs\"acker-Williams gluon distribution}  \label{sec:WW_Orho4}

To leading order in $A^+$ the field in L.C.\ gauge is given by
$A^i(\vec q) = -i q^i \, A^+(q)$. This leads to the WW gluon
distributions
\bea \label{eq:appC-xG-xh-LO}
\delta^{ij} \left< A^{ia}(\vec q)\, A^{ja}(-\vec q)\right> &=&
\left(2\frac{q^iq^j}{q^2}-\delta^{ij}\right)\,
\left< A^{ia}(\vec q)\, A^{ja}(-\vec q)\right> = 
\frac{N_c^2-1}{2q^2} \, g^2\, G_2(\vec q, -\vec q)~.
\eea
At this order the conventional and linearly polarized gluon
distributions are equal, and there is maximal polarization. Due to
``color neutrality'' of the proton, $G_2(\vec q, -\vec q) / q^2$ does
not diverge as $q\to0$.\\

\begin{figure}[htb]
  \centering
  \includegraphics[width=0.45\textwidth]{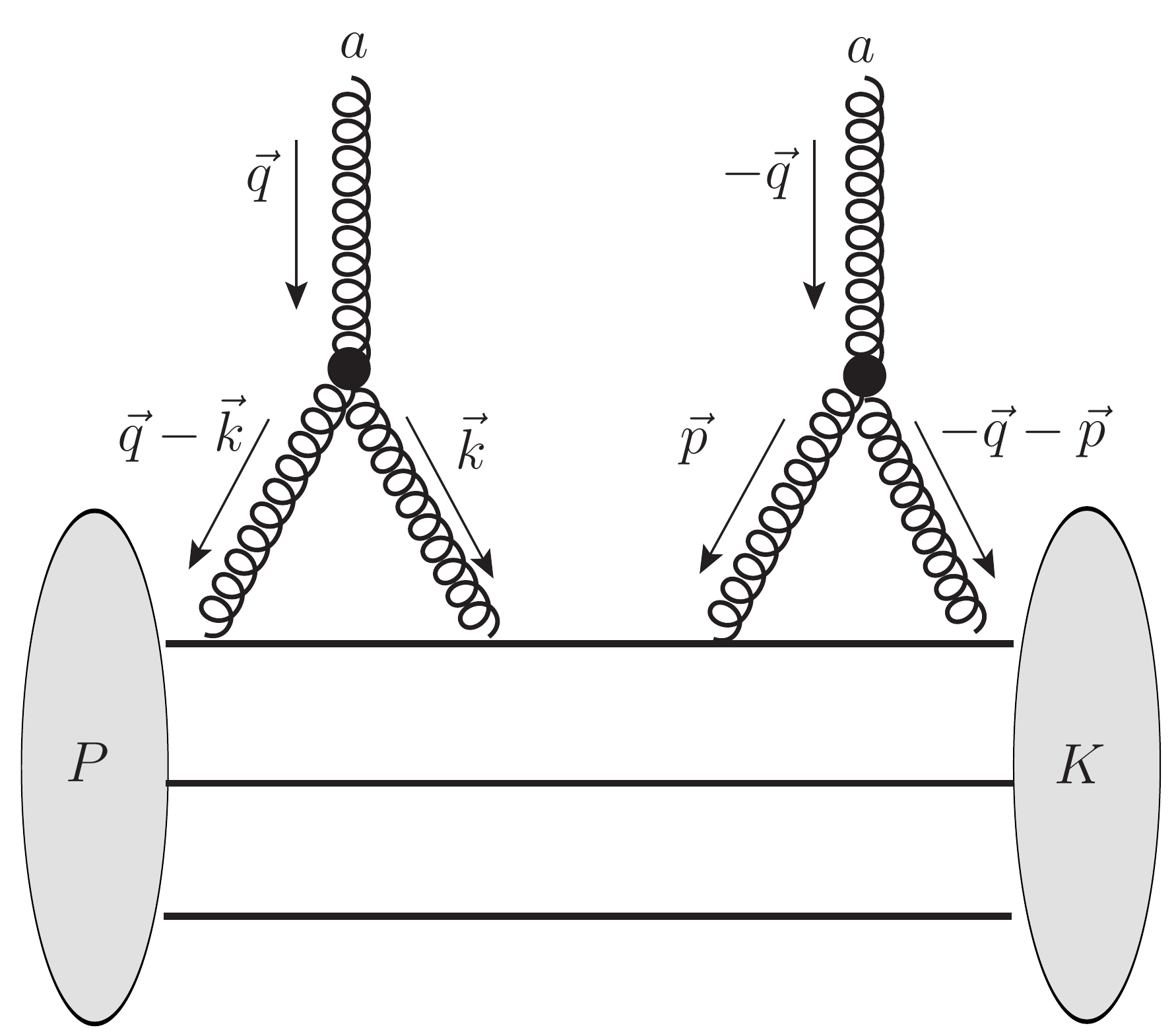}
  \caption{One of the diagrams for the WW gluon distribution at fourth
    order in $gA^+$.}
\label{fig:WW_O4}
\end{figure}
Solving eq.~(\ref{eq:cov--LC}) to quadratic order in $A^+$ one
has~\cite{Kovner:1995ts,Kovchegov:1997pc}
\be \label{eq:Ai_A+^2}
A^{ia}(\vec q) = -i q^i \, A^{+a}(\vec q) + \frac{ig}{2} f^{abc}
\left(\frac{q^iq^j}{q^2}-\delta^{ij}\right)
\int_k k^j A^{+b}(\vec q-\vec k) \, A^{+c}(\vec k)~.
\ee
This corresponds to the soft, ``quasi-classical'' field of recoil less
valence quark sources. It is assumed that the contribution from
diagrams corresponding to the internal exchange of a gluon over a large
longitudinal distance $x^-$ is suppressed, see the detailed discussion by
Kovchegov in ref.~\cite{Kovchegov:1997pc}.

The contribution to $A^i$ at quadratic order in $A^+$
leads to a correction to the WW gluon distributions at fourth
order in $A^+$ (fig.~\ref{fig:WW_O4})~\cite{Dumitru:2017ftq}
\bea
\Delta xG^{(1)}(x,\vec q) = - \Delta xh_\perp^{(1)}(x,\vec q) &=&
\frac{1}{4\pi^3} \frac{g^2}{4} f^{abe} f^{cde} \int_{k,p}
\left(\frac{\vec k\cdot \vec q \,
 \, \vec p\cdot \vec q}{q^2} - \vec k\cdot \vec p\right) \nn\\
& & \left<
A^{+a}(\vec q-\vec k)\, A^{+b}(\vec k)\, A^{+c}(-\vec q-\vec p)\,
A^{+d}(\vec p)
\right> \label{eq:xGxh_A+4}\\
&=&
\frac{1}{4\pi^3} \frac{g^2}{4} f^{abe} f^{cde} \int_{k,p}
\frac{1}{k^2}\frac{1}{p^2}\frac{1}{(\vec q-\vec k)^2}\frac{1}{(\vec q
  + \vec p)^2}
\left(\frac{\vec k\cdot \vec q \,\,
  \vec p\cdot \vec q}{q^2} - \vec k\cdot \vec p\right)
 \nn\\
& & \left<
\rho^a(\vec q-\vec k)\, \rho^b(\vec k)\, \rho^c(-\vec q-\vec p)\,
\rho^d(\vec p) \right>~.  \label{eq:xGxh_rho4}
\eea
There is no contribution from cubic order in $A^+$ as this is
proportional to the product of the longitudinal L.C.\ gauge field
$A^i$ at leading order with the transverse part of $A^j$ at quadratic
order (or vice versa), contracted with either $\delta^{ij}$ or
$\left(2\frac{q^iq^j}{q^2}-\delta^{ij}\right)$, which gives zero. Note
that the parenthesis in eqs.~(\ref{eq:xGxh_A+4}, \ref{eq:xGxh_rho4})
can also be written in terms of the 2d cross product as $[(\vec q-\vec
  k)\times \vec q] \, [(\vec q+\vec p)\times \vec q] / q^2$.\\

With $\langle\rho^4\rangle$ from eq.~(\ref{eq:rho4_full}) and
\bea
f(\vec q) &=& 
\int \dd x_1 \dd  x_2 \dd x_3 \, \delta(1-x_1-x_2-x_3)
\int \frac{\dd^2 p_1 \dd^2 p_2 \dd^2 p_3}{(16\pi^3)^2}
\, \delta(\vec{p}_1+\vec{p}_2+\vec{p}_3) \nn\\
& &
~~~~~\psi^*(\vec p_1 - \vec q, \vec p_2 + \vec q, \vec p_3) ~
\psi(\vec p_1, \vec p_2, \vec p_3) ~, \\
g(\vec q_1, \vec q_2) &=&
\int \dd x_1 \dd  x_2 \dd x_3 \, \delta(1-x_1-x_2-x_3)
\int \frac{\dd^2 p_1 \dd^2 p_2 \dd^2 p_3}{(16\pi^3)^2}
\, \delta(\vec{p}_1+\vec{p}_2+\vec{p}_3) \nn\\
& &
~~~~~\psi^*(\vec p_1 - \vec q_1, \vec p_2 - \vec q_2, \vec p_3+
\vec q_1 + \vec q_2) ~
\psi(\vec p_1, \vec p_2, \vec p_3) ~,
\eea
we can write the correction in the form
\bea
\Delta xG^{(1)}(x,\vec q) &=& - \Delta xh_\perp^{(1)}(x,\vec q) = \nn \\
& & \frac{1}{4\pi^3} \frac{3}{4}\, g^6
\int_{k,p}
\frac{1}{k^2}\frac{1}{p^2}\frac{1}{(\vec q-\vec k)^2}\frac{1}{(\vec q
  + \vec p)^2}
\left(\frac{\vec k\cdot \vec q \,
  \, \vec p\cdot \vec q}{q^2} - \vec k\cdot \vec p\right) \nn\\
& &
\left[-3 +3f(\vec q) +2 f(\vec p+\vec k) -2 f(\vec p+\vec q-\vec
  k) +3 f(\vec p) + 3 f(\vec q-\vec k) - 3 f(\vec p+\vec q)
  - 3 f(\vec k) \right.\nn\\
  & &\left.
  + g(\vec k, \vec p)- g(\vec p+\vec q
  - \vec k, \vec k)- g(\vec q - \vec k, \vec p) +
  g(\vec q - \vec k, \vec p+\vec k)
  \right]~.  \label{eq:dxG_dxh_fg}
\eea
The bracket vanishes if any two momenta ($\vec p, \vec q$ or $\vec k,
\vec q$ or $\vec p, \vec k$) are taken to zero. At finite $\vec q$ the
integral is free of infrared divergences and can be evaluated by
Monte-Carlo integration.\\

The correction at order $(A^+)^4$ increases with decreasing transverse
momentum and eventually overwhelms the leading contribution $\sim
(A^+)^2$. At such low $\vec q$ the result can no longer be trusted, and a
resummation to all powers of $A^+$ would be required. However, it is
interesting to note that at very small $x$ some configurations of the
proton correspond to negative $xh_\perp^{(1)}(x,q)$ at $q$ of order the
saturation scale, even when the function is resummed to all orders in
$A^+$~\cite{Dumitru:2017ftq}.\\~~\\

If the four charge correlator of eq.~(\ref{eq:xGxh_rho4}) is replaced
by a sum over pairwise contractions,
\bea
\left<\rho^a(\vec q-\vec k)\, \rho^b(\vec k)\, \rho^c(-\vec q-\vec p)\,
\rho^d(\vec p) \right> &\to&
\left<\rho^a(\vec q-\vec k)\, \rho^b(\vec k)\right>\,
\left<\rho^c(-\vec q-\vec p)\, \rho^d(\vec p) \right> +
\left<\rho^a(\vec q-\vec k)\, \rho^c(-\vec q-\vec p)\right>\,
\left<\rho^b(\vec k)\, \rho^d(\vec p) \right> \nn\\
&+&
\left<\rho^a(\vec q-\vec k)\,\,\rho^d(\vec p)\right>\,
\left< \rho^b(\vec k)\, \rho^c(-\vec q-\vec p)\right>
 ~,  \label{eq:rho4-rho2rho2}
\eea
then the correction to the WW gluon distribution becomes
\bea
\Delta xG^{(1)}(x,\vec q)\Big|_{\text{Gauss}} &=& - \Delta
xh_\perp^{(1)}(x,\vec q)\Big|_{\text{Gauss}} \nn\\
&=&
\frac{g^6\, N_c (N_c^2-1)}{32\pi^3}  \int_{k,p}
\frac{1}{k^2}\frac{1}{p^2}\frac{1}{(\vec q-\vec k)^2}\frac{1}{(\vec q
  + \vec p)^2}
\left(\frac{\vec k\cdot \vec q \,\,
  \vec p\cdot \vec q}{q^2} - \vec k\cdot \vec p\right)
\, G_2(\vec k-\vec q,\vec q+\vec p) \,\, G_2(\vec k,\vec p)
 ~.  \label{eq:xGxh_rho4_Gauss}
\eea
Note that the $\langle\rho^2\rangle$ correlators
in~(\ref{eq:rho4-rho2rho2}) are non-forward matrix elements. The
dominant contribution to the integral in
eq.~(\ref{eq:xGxh_rho4_Gauss}) is from $|\vec k + \vec p|$ on the
order of the transverse momentum of the quarks in the proton so that
both $G_2$ correlators are evaluated for small momentum transfer;
their 1-body GPD limit suffices for high $q$.

\begin{figure}[htb]
  \centering
  \includegraphics[width=0.45\textwidth]{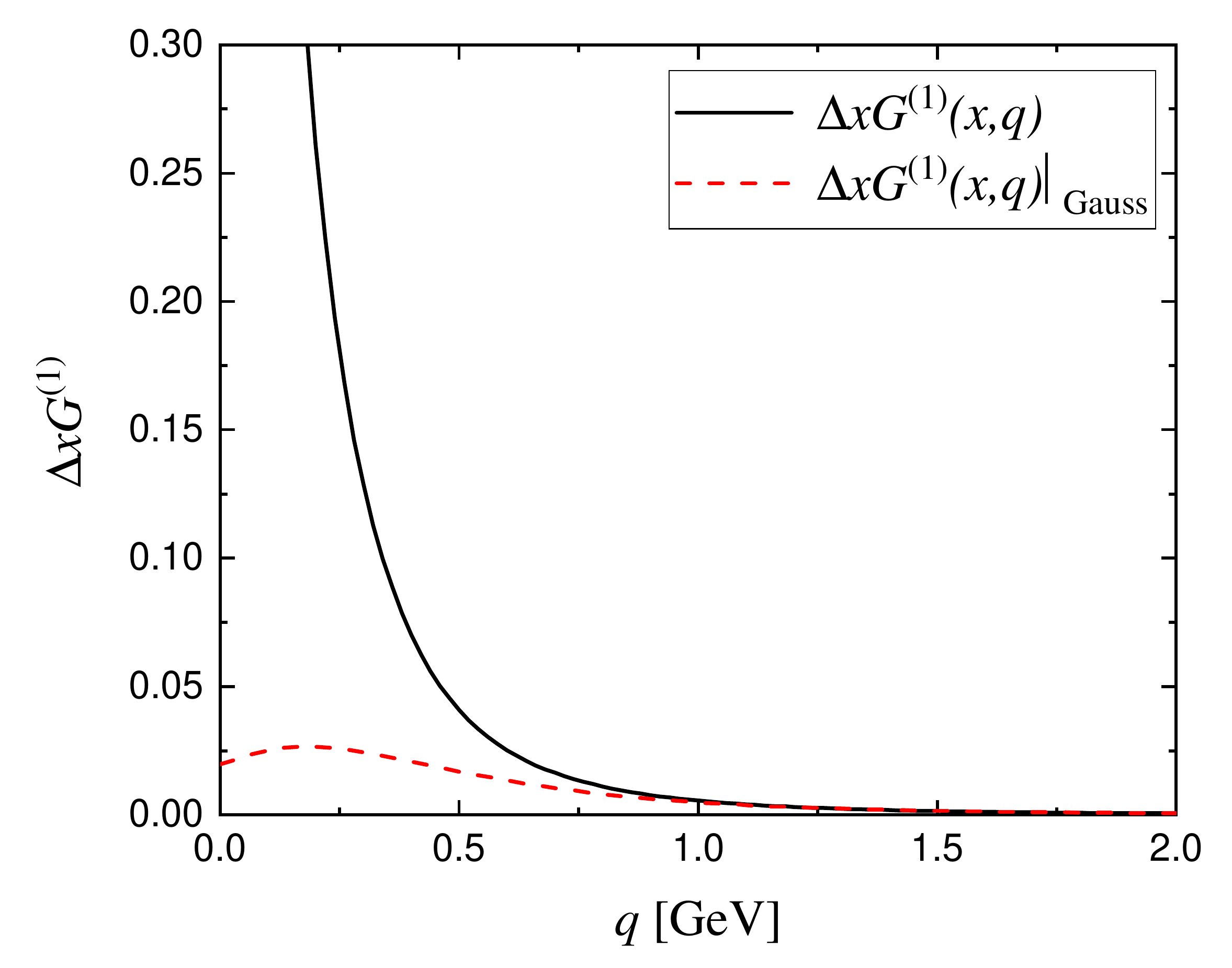}
  \caption{The next-to-leading twist correction to the WW gluon
    distributions. The complete correlator of four $A^+$ is compared
    to the sum over pairwise quadratic correlators.}
\label{fig:WW_delta-xG_vs_Gauss}
\end{figure}

Fig.~\ref{fig:WW_delta-xG_vs_Gauss} shows a numerical comparison of
eq.~(\ref{eq:xGxh_rho4_Gauss}) to the complete
result~(\ref{eq:dxG_dxh_fg}). They agree at high transverse momentum
where, however, the correction due to the transverse part of $A^{ia}$
is much smaller than the leading contribution. At $q\sim 0.2$~GeV the
Gaussian approximation we described underestimates the true correction
to the WW gluon distributions by about one order of magnitude.
\\

Finally, we present expressions for the resummed form of
$xG^{(1)}(x,\vec q)$ and $xh_\perp^{(1)}(x,\vec q)$ in a (large-$N_c$)
Gaussian approximation for the general correlator
$\langle \rho^a(\vec q_1)\, \rho^b(\vec q_2) \rangle = \frac{1}{2}\,
\delta^{ab}\, g^2\, G_2(\vec q_1, \vec q_2)$. Relaxing the assumption
of translational invariance in the transverse plane, eqs.~(30, 31) of
ref.~\cite{Dumitru:2016jku} become
\begin{eqnarray}  \label{eq:appC-xh-Gauss}
x h_\perp^{(1)} (x, \vec{q})  &=& \frac{N_c} {\alpha_s} \int \frac{\dd^2
  r}{(2\pi)^2} \int \frac{\dd^2 b}{(2\pi)^2}\,
e^{- i \vec{q} \cdot \vec{r}} (1-S^2) \frac{1}{\Gamma} \left(
2 (\hat{q} \cdot \vec{\nabla}_r)^2-\nabla_r^2  \right) \Gamma\\
x G^{(1)} (x, \vec{q})  &=& \frac{N_c} {\alpha_s} \int \frac{\dd^2
  r}{(2\pi)^2} \int \frac{\dd^2 b}{(2\pi)^2}\,
e^{- i \vec{q} \cdot \vec{r}} (1-S^2) \frac{1}{\Gamma} \nabla_r^2\, \Gamma
~.  \label{eq:appC-xG-Gauss}
\end{eqnarray}
Here,
\be \label{eq:appC-S}
S(\vec r, \vec b) = \exp\left( -\frac1{2} C_F \Gamma(\vec r,
\vec b)\right) ~,
\ee
denotes the dipole scattering matrix, and
\be \label{eq:appC-Gamma}
\Gamma(\vec r, \vec b) = (4\pi\alpha_s)^2 \int\limits_{p,q}
\frac{e^{i (\vec p -\vec q)\cdot \vec b}}{p^2 q^2}\left(1 - e^{i (\vec
    p + \vec q)\cdot \frac{\vec r}{2}}\right)\, G_2(\vec p, -\vec q)~.
\ee
The MV model correlator is recovered if one averages $\vec b$ over a
large transverse area $S_\perp$, and replaces $G_2(\vec p, -\vec p)$
by a constant proportional to $\mu^2 S_\perp$ (which also requires one
to introduce an IR cutoff $\Lambda_{\rm IR}$).

We refrain from a numerical
evaluation of eqs.~(\ref{eq:appC-xh-Gauss} - \ref{eq:appC-Gamma}) here
which is rather tedious. Given that the saturation scale for
non-linear dynamics in the proton at $x\sim 0.1$ is rather small, we
expect that for $q\simge 0.5$~GeV the resummation does not give a
significant correction to eq.~(\ref{eq:appC-xG-xh-LO}) either.


\end{document}